# pH/T duality – wall properties and time evolution of plant cells

Mariusz A. Pietruszka

Faculty of Biology and Environment Protection, Plant Physiology, University of Silesia, Katowice, Poland

E-mail: mariusz.pietruszka@us.edu.pl

**Abstract** We examined the pH/T duality of acidic pH and temperature (T) for the growth of grass shoots in order to determine the phenomenological 'equation of state' (EoS) for living plants. By considering non-meristematic growth as a dynamic series of 'state transitions' (STs) in the extending primary wall, we identified the critical (read: optimum) exponents for this phenomenon, which exhibit a singular behaviour at a critical temperature, critical pH and critical chemical potential (μ) in the form of four power laws for 'state functions' (SFs): $F_\pi(\tau) \propto |\tau|^{\beta-1}$, $F_\tau(\pi) \propto |\pi|^{1-\alpha}$, $G_\mu(\tau) \propto |\tau|^{-2-\alpha+2\beta}$ and $G_\tau(\mu) \propto |\mu|^{2-\alpha}$. The power-law exponents α and β are numbers, which are independent of pH (or μ) and T that are known as critical exponents, while π and τ represent a reduced pH and reduced temperature, respectively. Various scaling predictions were obtained – a convexity relation α + β ≥ 2 for practical pH-based analysis and a β ≡ 2 identity in microscopic representation. In the presented scenario, the magnitude that is decisive is the chemical potential of $H^+$ ions (protons), enforcing subsequent STs and growth. The EoS span areas of the biological, physical, chemical and Earth sciences cross the borders with the language (adapted formalism) of phase transitions. Furthermore, we observed that the growth rate is generally proportional to the product of the Euler beta functions of T and pH, which are used to determine the hidden content of the Lockhart constant Φ. It turned out that the evolution equation, when expressed in terms of the identical (pH, T), dynamic set of variables, explains either the usually observed (averaged) monotonic growth or periodic extension – like the one detected in pollen tubes – in a unified account. The pH drops induced by auxin or fusicoccin, when next converted by the augmented Lockhart equation, are enough to explain a significant fraction of the increase in growth rate with exceptionally high determination coefficients ($R^2 \approx 0.99997$). The self-consistent recurring model is proposed to embrace the inherent complexity of such a biological system, in which several intricate pathways work simultaneously, in order to reconcile the conflicting views of plant cell expansion and growth.

**Author summary**

In plant development, sudden changes such as cell expansion or pollen tube oscillations seem to depend on a correlative group of events rather than slow shifts in the apex. Hence, in order to understand or to control the processes in the expanding cell wall, we need to unravel the general principles and constraints that govern growth. The quest for these principles has primarily focused on the molecular, though merely descriptive, level. Here, we show that it is possible to analyse oscillatory phase changes computationally without even requiring knowledge about the exact type of transition. Our results suggest that the cell wall properties and growth of plant cells can be accurately and efficiently predicted by a set of physical and chemical variables such as temperature, pressure and the pH of the growing plant, which build a scaffold for more specific biochemical predictions. The model equations that we propose span areas of the biological, physical, chemical and Earth sciences. A common denominator that ties the growth factors together is the chemical potential of protons.

# PART I

# 'Equation of State' for Plants

**Introduction**

*General outline*

This part deals with the derivation and determination of an 'equation of state' (EoS) for plants that integrates the relationship between temperature and pH (or chemical potential, µ) with growth. Based on reliable (published) data and our own experimental data, we sought to identify the 'state' variables that are necessary for optimal growth. The limited number of parameters can be misleading; this approach does not oversimplify the complexity of biological systems, but conversely, takes all of them into account by building an 'outer scaffold' that cannot be surpassed. We cannot dismiss the possibility that this is the role of phenomenology in making universal statements, in this case imposing a physical constraint (i.e. EoS) on the extending primary wall.

A commonly held view is that biological systems are complex (and molecular biologists are deluged with data), and that therefore modelling has to cope with the fact that the degrees of freedom are numerous, and correspondingly, the number of parameters should be high. Though physical systems on a microscopic scale are also complex and the number of degrees of freedom in real systems is abundant (e.g., Avogadro number, as in the solid state physics), their behaviour can usually be described by a scarce number of parameters (Ginzburg, 2004) within the framework of a phenomenological theory. A similar argument also concerns quantum models (Hubbard, 1963), which has only two parameters: t – for the "hopping" integral and U – for the Coulomb interaction of electrons within narrow energy bands, or the t-J model, which was first derived in 1977 from the Hubbard model by Spalek (2007) – also with a two-parameter (t, J) space. In what follows, I argue that, similar to physical systems, a low number of relevant parameters may also be sufficient in complex biological systems. To support this view let me quote Portes et al. (2015): "This is not to invalidate the strive for detailed models, which are necessary to understand the role of specific components of interest, however, they require a compatible amount of data. In contrast, more general models help to unveil the fundamental principles driving a phenomenon of interest, helping to identify the key regulatory interactions, albeit lacking details". In what follows we have chosen the second route.

The growth of plant cells and organs (like extending cylindrical organs such as grass shoots) are affected by light and humidity, which along with pH and temperature are the most important factors that influence growth. It can be a rewarding enterprise to understand plant growth in terms of the time evolution of STs (e.g., from ordered to disordered state, from stressed to relaxed state, from "covalent bonds state" to "disruption of covalent bonds state", etc.) that take place in the peripheral cell wall. Such transition (ST, which should <u>not</u> be confused with a phase transition encountered in physics) occur *via* the exchange of particles and energy-consuming metabolic processes (such as energy-conserving ATP) with the inside of cell compartment – the vacuole and cytoplasm, which are treated as the reservoir of molecules and heat (thermostat). Hence, we consider living organisms as *open systems* that dissipate energy and exchange entropy (heat) and matter (molecules) with the environment (Barbacci et al., 2015). This issue is accomplished by the presence of pH and T dependent Euler beta functions in model equations representing the modes of interaction (like heat and mass transfer) with the outside world of a growing plant. Even though we considered single ST, a cascade (ratchet) of subsequent chemical potential – induced STs, which constitute the extension of the cell wall and growth, can be imagined (Part II). In what follows we suggest that well established phase transition techniques can be successfully engaged in the description of optimum growth. We demonstrate the prevailing role of the chemical potential and propose that growth of plant cells and organs is regulated, at the molecular level, by the chemical potential (Matlak et al., 2004) of $H^+$ ions. The proposed model correctly accounted for the set of experimental results and predicted the growth of grass shoots from pH with unprecedented fidelity.

*Preliminaries*

Plant developmental systems have evolved within the universal limitations that are imposed by the plant cell wall (Lintilhac, 2014). Modulation of the mechanical properties occurs through the control of the biochemical composition and the degree and nature of interlinking between cell wall polysaccharides (Bidhendi and Geitmann, 2015). Plant cells encase themselves in a complex polysaccharide wall (Cosgrove, 2005a), and characteristically obtain most of their energy from sunlight *via* the photosynthesis of the primary chloroplasts. The expansive growth of turgid cells, which is defined as an irreversible increase in cell volume, can be regarded as a physical process that is governed by the mechanical properties of the cell wall and the osmotic properties of the protoplast (Schopfer, 2006). The precise biochemical mechanism that regulates the ability of the growth-limiting walls to extend irreversibly under the force of turgor pressure has not yet been identified (Kutschera, 2000), but it is correlated with loosening of cells walls, what increase the cell susceptibility for expansion.

Growing plant cells characteristically exhibit acid growth (Rayle and Cleland, 1970; Hager et al., 1971; Cosgrove, 1989), which has been formulated in the form of the "acid growth hypothesis" (Hager, 2003; Taiz and Zeiger, 2006). The acid growth hypothesis was proposed over 40 years ago and is widely considered to play a role in expansive cell growth. The acid growth hypothesis postulates that both phytotoxin fusicoccin (FC) and the growth-promoting factor auxin (indole-3-acetic acid, IAA) cause wall-loosening and produce the concomitant induction of growth (growth enhancement) through the rapid acidification of the extension-limiting cell wall (Cleland, 1973; Kutschera, 1994). These enigmatic "wall-loosening processes" are in fact minor changes within the polymer network of the extension limiting walls (i.e., the incorporation of proteins, the enzymatic splitting of polymer backbones or covalent cross-links or the disruption of non-covalent interactions between wall polymers *via* expansins activity, see pioneering study of McQueen-Mason and Cosgrove, 1994). Molecular studies revealed that expansins are a large family of proteins with 38 members in *Arabidopsis thaliana*, which might be divided into three subfamilies: α-, β- and γ-expansins (Li et al. 2002). It was shown that individual expansins might be tissues- or process-specific, and loss-of-function mutation resulted in the disorder in specific aspect of plant growth or development, for example in root hair development (reviewed by Marzec et al., 2015a).

Growth is accomplished through the enlargement of the cell volume owing to water uptake, which maintains the appropriate inside pressure in the vacuole as well as the irreversible extension of the pre-existing (primary) cell wall. Hereafter, we will assume an almost constant or slowly varying turgor pressure, which by definition is a force that is generated by water pushing outward on the plasma membrane and plant cell wall that results in rigidity in the plant cell and the plant as a whole.

Expansive growth is the result of the coupling effects (Barbacci et al., 2013) between mechanical (pressure), thermal (temperature) and chemical energy (pH). The thermal sensitivity of biochemical processes refers to the coupling effect between the thermal and chemical energies. In the search for a plant-wall specific EoS in growing biological cells or non-meristematic tissues, we considered two 'state variables', namely temperature and pH at a constant turgor. We note that pH is not *sensu stricto* a fundamental physical quantity, and cannot be treated as a usual intense variable. Nonetheless, relying on the definition of pH and then considering the chemical potential (Baierlein, 2001), we may propose the following.

Evidence has accumulated that the final goal of auxin action (Steinacher et al., 2012; Lüthen, 2015 for review) is to activate the plasma membrane (PM) $H^+$-ATPase, a process which excretes $H^+$ ions into the cell wall. The auxin-enhanced $H^+$-pumping lowers the cell wall pH, activates pH-sensitive enzymes and proteins, such as expansins (Cosgrove, 1993), xyloglucans (Fry et al., 1992) or yieldins (Okamoto-Nakazato et al., 2001), within the wall and initiates cell-wall loosening and extension growth (Hager, 2003). It has also been observed that when auxin-depleted segments were submerged in an acid buffer solution, the segments started their elongation growth immediately ("acid growth"), whereas the auxin-induced growth ("auxin growth") began after a delay (lag phase), as is shown in Fig. 1 (curve 1) in Hager (2003); see also the remaining plots, which are of quite a different character, (curves 2 – 6) that correspond to "acid growth". From the many investigations that have been done for more than four decades, it was deduced that protons, which are exerted into the wall compartment, are directly responsible for wall-loosening processes through the hydrolysis of covalent bonds, transglycosylation or the disruption of non-covalent bonds (changing the 'state' of the system). The growth effect is illustrated in Fig. 2 in Hager (2003), where the "Zuwachs" (increment in %) is plotted against pH.

However, pH can be expressed in terms of the more elementary quantity, namely the chemical potential $\mu_{H+}$ of protons ($H^+$). Note that indirect measurements of the chemical potential by means of pH can be compared to our previous measurements of the electromotive force (EMF) in order to detect phase transitions, which are localised by the 'kinks' in the chemical potential in condensed matter physics (Matlak and Pietruszka, 2000; Matlak et al. 2001). Although pH is not a typical generalised coordinate, the microscopic state of the system can be expressed through it in a collective way, similar to the recent measurement of the EMF induced by oscillating ionic fluxes in the (tip) growing plants (Pietruszka and Haduch-Sendecka, 2015a).

Temperature is among the most important environmental factors that determine plant growth and cell wall yielding (Pietruszka et al., 2007). Plants are responsive to temperature, and some species can distinguish differences of 1°C. In Arabidopsis, warmer temperature accelerates flowering and increases elongation growth (Jung et al., 2016). In many cases the process of growth can be differentiated by the response of plants to temperature (Went, 1953). Actually, only a few papers in which temperature response is treated as a major problem can be mentioned (Yan and Hunt, 1999 and papers cited therein), but some of recent publications highlighted the role of temperature, as a factor that regulated the plant development, i.e. *via* phytohormones (Dockter et al., 2014, Jung et al., 2016).

Now, inspired by the Ansatz using Euler beta function f(x, α, β) for growth empirical data, we will show that the final action of temperature and pH on plant growth is effectively the same, though both triggers of these responses are apparently of a different nature.

**Material and methods**

*Plant material*

Experiments were done using three-day-old seedlings of *Zea mays* L cv. Cosmo. Seeds of maize were soaked in tap water for 2h, and then sown in moist lignin. The seedlings were grown in an incubator in darkness at 27 ± 0.5 °C for three days. Ten mm segments were cut from the plants 5 mm below the tip and first leaf was removed.

*Measurement of growth and acidification*

Experiments were performed for two cases: for artificial pond water (APW) and fusicoccin (FC) at a concentration of $10^{-6}$ M. The growth stimulus FC was introduced into the incubation medium after 2 hours. Thirty ten-mm-long coleoptile segments were prepared for each variant and then placed in two identical sets (15) of the elongation-measuring apparatus in an aerated incubation medium. The volume of the incubation medium in the apparatus was 5 ml per glass tube (0.3 ml of the incubation medium per segment). Measurements of the growth and pH were performed for 12.5 hours and recorded every 15 minutes. The pH measurement was carried out using CPI-501 pH-meter. Images of the segments were recorded using a CCD camera (Hama Webcam AC-150). Only one of the two tests in separate tubes was selected for analysis. The length of the coleoptile segments (initially 10-mm-long segments) was measured in ImageJ program with the accuracy established at the ± 0.1 mm level. The relative elongation was calculated using the formula $(l_t - l_0)/l_0$ with l for "length" and $l_0 = l(t = 0)$. Then the relative volume was obtained by multiplying length by the average cuticle segment cross-section area. The temperature during the experiment was 25 ± 0.5°C and was maintained using a water bath at a similar level. All manipulations were performed under dim green light.

*Construction of a phenomenological model*

Based on the "acid growth hypothesis" and relevant experimental data (Yan and Hunt, 1999; Hager, 2003), we examined the pH/T (or μ/T) duality of acidic pH (or auxin-induced acidification) and temperature (T) for the growth of grass shoots, in order to determine the EoS for extending the primary wall of (living) plants.

pH is defined as the decimal logarithm of the reciprocal of the hydrogen ion activity $a_{H+}$ in a solution (pH = $-\log_{10} a_{H+}$ = $\log_{10} 1/a_{H+}$). The pH value is a logarithmic measure of $H^+$-activity (the tendency of a solution to take $H^+$) in aqueous solutions and defines their acidity or alkalinity; $a_{H+}$ denotes the activity of hydronium ions in units of mol/l. The logarithmic pH scale ranges from 0 to 14 (neutral water has a pH equal to 7). The pertinent experimental data are adapted from Hager (2003).

Temperatures at which most physiological processes occur normally in plants range from approximately 0°C to 45°C, which determines the physiological temperature scale (in Kelvin scale: [K] = [°C] + 273.15). The temperature responses of plants include all of the biological processes throughout the biochemical reactions (high $Q_{10}$ factor). This is clearly visible in the relative rates of all of the development or growth processes of maize as a function of temperature, which is illustrated in Fig. 3 in Yan and Hunt (1999). For future applications, we propose that the 'absolute' temperature scale for plants [0 – 45] °C, which corresponds to a [0, 1] interval after rescaling, be setup. Note, this is *not* a real absolute temperature.

Which data have been adopted for the research and how those data have been handled is described beneath. To be precise, for the analysis, we accepted the reliable published data, namely the pH plots that are presented in Fig. 2 in Hager (2003) and the temperature plots in Fig. 3 in Yan and Hunt (1999). These charts were first digitised by GetData digitiser for reconstruction and then rescaled (normalised to [0, 1] interval) for both perpendicular axes (SI Figure 1), dividing by a maximum value. For example, the original temperature range [0 – 45] °C was converted to [0, 1] interval through division by 45. The same procedure was conducted on pH scale [0, 14] to obtain the normalized [0, 1] interval (SI Figure 1), indispensable for further calculations.

The pH plots that are presented in Fig. 2 in Hager (2003) as well as the temperature plots in Fig. 3 in Yan and Hunt (1999) look similar when mirror symmetry is applied. Providing that we perform a substitution x → (1 – x) and consider a function f(x) of a variable x and of its reflection (1 – x), one might expect that they will look approximately the same after proper scaling. This properties can be adequately described by the Euler beta density distribution (or beta Euler, for short). In this approach, we assume that x equals either pH or temperature (both rescaled to a [0, 1] interval), an assignment which makes sense, since at relatively low pH, which corresponds to a relatively high T in this [0, 1] scaling, plant cells and organs grow (elongate) the fastest, which is clearly reproduced in SI Figure 1.

The data from SI Figure 1a were taken separately ("auxin" and "acid" growth; Hager, 2003) for the fitting procedure (SI Table 1 and SI Figure 2), while the 'temperature' data (Yan and Hunt, 1999) from SI Figure 1b were first normalized and then fitted by beta Euler.

Furthermore, when comparing data from different experiments (Yan and Hunt, 1999; Hager, 2003), we observed that the x variable may, after rescaling, bear the following meaning – either acidity (basicity) x ≡ pH or temperature x ≡ T (see SI Figure 1) with a characteristic beta function cut-off at x = 0 (SI Figure 1a and the inset) and at x = 1 (SI Figure 1b); the normalised scale we used is presented in SI Figure 1b (inset).

**Results**

*Derivation of the EoS for plants*

The probability density function of the beta distribution (also called the Euler integral of the first kind (Polyanin and Chernoutsan, 2011) for 0 ≤ x ≤ 1 and shape parameters α, β > 0 is a power function of the variable x

$$f(x;\alpha,\beta) = \text{const} \cdot x^{\alpha-1}(1-x)^{\beta-1} = \frac{x^{\alpha-1}(1-x)^{\beta-1}}{\int_0^1 u^{\alpha-1}(1-u)^{\beta-1} du} = \frac{1}{B(\alpha,\beta)} x^{\alpha-1}(1-x)^{\beta-1} \tag{I-1}$$

where B = B(α, β) is the normalisation constant. Based on the experimental observations as discussed in the previous sections and presented in SI Figure 1, the left side of equation (I-1), rewritten twice for x = pH or x = T, can be merged into a single expression

$$[c_{\text{pH}} \cdot \text{pH}^{\alpha-1}(1-\text{pH})^{\beta-1}]_T = [c_T \cdot T^{\beta-1}(1-T)^{\alpha-1}]_{\text{pH}} \tag{I-2}$$

where $c_{\text{pH}}$ and $c_T$ are constants and "pH" is treated here as a non-separable variable name. The lower indices "T" and "pH" in Eq. (I-2) should be read "at constant temperature", and "at constant pH", respectively, which is valid for preparation of strict experimental conditions (the use of a pH-stat apparatus as in Lüthen et al., 1990, to achieve a stable equilibrium pH seems to be indispensable; see also Hager, 2003). Equation (I-2) will be considered further as a constraint on two independent variables T and pH.

By defining the SF *F* at constant turgor pressure *P* (isobaric conditions: δP ≈ 0)

$$F(T,\text{pH};\alpha,\beta)\big|_P \equiv \left(\frac{T}{1-\text{pH}}\right)^{\beta-1}\left(\frac{1-T}{\text{pH}}\right)^{\alpha-1} \tag{I-3}$$

that is an analytic function of pH and T on the physically accessible interval [0, 1) for each variable (note that *F* is <u>not</u> the free energy of the system; here *F* constructs exclusively a bridge between physical and chemical features of the living system), and assuming adequate water uptake to fulfil this requirement, we determine from equation (I-2)

$$\frac{T^{\beta-1}(1-T)^{\alpha-1}}{pH^{\alpha-1}(1-pH)^{\beta-1}} = \frac{c_{pH}}{c_T} \equiv \Omega_S \tag{I-4}$$

where $\Omega_S$ is a dimensionless constant. We notice that Eq. (I-4) takes, after transformation, the usual form of an equation of state. Hence, the constitutive EoS for living plant cells explicitly reads

$$\boxed{\left(\frac{T}{1-pH}\right)^{\beta-1}\left(\frac{1-T}{pH}\right)^{\alpha-1} = \frac{c_{pH}}{c_T}} \tag{I-5}$$

which has the elegant structure of a double power law. Note that T is the temperature in Celsius rescaled to a [0, 1] interval, as pH also is. Here α and β are the shape exponents (SI Table 1 and SI Figure 2). The saddle-point type of solution that is presented in Fig. 1 represents a concave surface in a form of a contour plot, which is roughly accurate at low pH and moderate temperatures. Equation (I-5) becomes increasingly inaccurate (divergent) at very low or very high pH values (close to zero or one in the normalised scale, which we use throughout the article), though such excessive conditions, which constitute unacceptable extremes for life to come to existence, are excluded by Nature. The so-called "threshold pH", was already observed by Lüthen et al. (1990). The isotherms of Eq. (I-3) are presented in Fig. 2a, while the lines of constant pH are presented in Fig. 2b. Note that the use of the term "equation of state" can be questionable, since it does not represent the relation between pH and T in the system from thermodynamic point of view. Therefore we use EoS instead.

Even though we (apparently) launched our derivation from the "acid growth theory", a closer look at the 'initial conditions' revealed that, in fact, it was established on the raw experimental data of pH and T dependent growth (SI Figure 1). For that reason, Eq. (I-5) is independent of whether the acid growth theory applies or not, and therefore seems universal, at least for tip-growing grass shoots. Note, Eq. (I-5) is not an evolution (growth) equation, but a system (cell wall) property equation. The solutions of Eq. (I-5) for α and β in the form of contour plots are presented in Figure 3.

Essential thermodynamic relations are property relations and for that reason they are independent of the type of process. In other words, they are valid for any substance (here: the primary wall of a given plant species) that goes through any process (mode of extension). Yet, in our case, we should be aware of the empirical origin of Eq. (I-5) and memorise the fact that a fundamental variable of the system is the chemical potential μ at a constant pressure and temperature (here: μ = $\mu_{H+}$(T) for $H^+$ ions). By taking the total derivative, where the implicit $\mu_{H+}$(T) dependence was taken into account, we get

$$dF(T, pH; \alpha, \beta) \approx dG(T, pH(T); \alpha, \beta) = dG(T, \mu_{H^+}(T); \alpha, \beta) \tag{I-6a}$$

where

$$G(T, \mu_{H^+}(T); \alpha, \beta)\big|_P \equiv -\frac{(\log 10)^\alpha}{T-1}(\mu_0 - \mu_{H^+})^2 \left[\frac{T-1}{(\mu_0 - \mu_{H^+})RT}\right]^\alpha \left[\frac{RT^2}{RT + (\mu_0 - \mu_{H^+})\log 10}\right]^{\beta-1} \quad \text{(I-6b)}$$

$G_P(T, \mu)$ is another SF (which should *not* be confused with the Gibbs energy) though expressed by intense, microscopic variables; temperature T here belongs to the [0, 1) interval. In approximation, since pH is measured in isothermal conditions (SI Figure 1a), we may temporarily suspend the pH(T) dependence in equation (I-6a) and consider pH and T as intense (non-additive) variables. Note, that Eqs (I-3) and (I-6) do not conclude about the free energy of the cell wall – they only describe mathematical relations between pH, T and $\mu_{H+}$ variables.

In physics, the equation of state is the relation between state variables that describes the state of matter under a given set of physical conditions. It is a constitutive equation that provides a relationship between two or more state functions that are associated with the matter, such as its temperature, pressure, volume or internal energy. They are useful in describing the properties of gases (ideal gas law: PV = nRT), fluids (van der Waals equation of state), condensed matter and even the interior of stars or (at present) the accelerated expansion of the Universe: P = wρ (Weinberg, 1972). In this context, equation (I-5) is the EoS that describes plant cell (wall) properties, where the SFs are the scaled temperature and pH of the extending cell volume. Here pH, which is intimately connected with the acid growth hypothesis, even though it is not an intense variable in the usual sense, introduces a direct link between the physical variables (P, V) that represent the state of the system (growing cell or tissue) and biological response (*via* pH altering – $H^+$-ATPase), which results in growth. Clearly, equation (I-5) can be further verified by the diverse experiments that have been conducted for many species in order to determine characteristic triads (α, β, $\Omega_S$) that belong to different species or families (classes) taxonomically, or that are simply controlled by growth factors (growth stimulators such as auxin, fusicoccin or inhibitors such as $CdCl_2$). However, the real test of the utility of Eq. (I-5) would include predictive outputs for any perturbations that could then be subject to experimental validation. From these triads, Eq. (I-5) should also be possible to connect with either the underlying physiological processes or their molecular drivers. In particular, whether the underlying molecular mechanism is identical or different should be reflected in the critical exponents (α and β pair should be identical for the same molecular mechanism and dissimilar for diverse mechanism; see also Eq. II-7 for a fitting procedure). Despite the fact that F is not a generalised homogeneous function, it can also be used to check for a kind of "universality hypothesis" (Stanley, 1971) and to validate the evolutionary paradigm with genuine numbers.

*Calculation of critical exponents*

Phase transitions in physics occur in the thermodynamic limit at a certain temperature, which is called the critical temperature $T_c$, where the whole system is correlated (radius of coherence $\xi$ becomes infinite at $T = T_c$). We want to describe the behaviour of the function F(pH, T), which is expressed in terms of a double power law by equation (I-3) that is close to the critical (optimum) temperature and, specifically in our case for practical reasons, also about the critical (optimum) pH = $pH_c$. In physics, critical exponents describe the behaviour of physical quantities near continuous phase transitions. It is believed, though not proven, that they are universal, i.e. they do not depend on the details of the system. Since growth (such as cell wall extension or the elongation growth of the shoots of grasses, coleoptiles or, hypocotyls) may be imagined as advancing over the course of time a series of a quasi-continuous STs (see Discussion), the critical exponents may be calculated. Analytical analysis must lead to such predictions that can be used as a Litmus test in order to cut through the myriad of conceptual models published in the biophysical community. In plant science, the calculation of critical exponents can be connected with the optimum growth. It can be helpful to identify of a molecular mechanism responsible for growth (similar exponents will be connected with the same mechanism).

Let us introduce the dimensionless control parameter ($\tau$) for the reduced temperature

$$\tau = 1 - \frac{T}{T_c} \tag{I-7}$$

and similarly ($\pi$), for the reduced pH

$$\pi = 1 - \frac{pH}{pH_c} \tag{I-8}$$

which are both zero at the transition, and calculate the critical exponents. It is important to keep in mind that critical exponents represent the asymptotic behaviour at phase transition, thus offering unique information about how the system (here: the polymer built cell wall) approaches a critical point. A critical point is defined as a point at which $\xi = \infty$ so in this sense $T = T_c$ and $pH = pH_c$ are critical points of Eq. (I-5). We believe that this significant property can help to determine the peculiar microscopic mechanism(s) that allows for wall extension (mode of extension) and growth in the future research. The information that is gained from this asymptotic behaviour may advance our present knowledge of these processes and their mechanisms and help to establish the prevailing one.

By substituting equations (I-7) and (I-8) into equation (I-3), we can calculate the critical exponents (when $T \to T_c$ and $\text{pH} \to \text{pH}_c$) from definition (Stanley, 1971), which will give us

$$\lambda_1 = \lim_{\tau \to 0} \frac{\log F(\tau, \pi)}{\log(\tau)} = \beta - 1 \tag{I-9}$$

$$\lambda_2 = \lim_{\pi \to 0} \frac{\log F(\tau, \pi)}{\log(\pi)} = 1 - \alpha \tag{I-10}$$

and

$$\lambda_3 = \lim_{\tau \to 0} \lim_{\pi \to 0} \frac{\log F(\tau, \pi)}{\log(\tau) \log(\pi)} = 0 \tag{I-11}$$

Here $\lambda_1 = \beta - 1$ and $\lambda_2 = 1 - \alpha$ are the critical exponents ($\lambda_3 = 0$). The above equations result in two power relations that are valid in the immediate vicinity of the critical points

$$F_\pi(\tau) \propto |\tau|^{\beta-1} \tag{I-12}$$

$$F_\tau(\pi) \propto |\pi|^{1-\alpha} \tag{I-13}$$

where lower indices denote constant magnitudes. Equations (I-12) and (I-13) represent the asymptotic behaviour of the function F(τ) as $\tau \to 0$ or F(π) as $\pi \to 0$. In fact, we can observe singular behaviour at T = $T_c$ (Fig. 4a) and pH = $\text{pH}_c$ (Fig. 4b). At this bi-critical point, the following convexity relation holds: α + β ≥ 2.

Retaining τ for the reduced temperature, let us introduce another – microscopic – control parameter, μ, for the chemical potential

$$\mu = 1 - \frac{\mu_{\text{H}^+}}{\mu_c} \tag{I-14}$$

Substituting τ and equation (I-14) into equation (I-6b) for G(T, μ) and assuming the reference potential $\mu_0$ = 0, we can calculate the critical exponents (when $T \to T_c$ and $\mu_{\text{H}^+} \to \mu_c$) to get

$$\lambda_1 = \lim_{\tau \to 0} \frac{\log G(\tau, \mu)}{\log(\tau)} = -2 - \alpha + 2\beta \tag{I-15}$$

$$\lambda_2 = \lim_{\mu \to 0} \frac{\log G(\tau, \mu)}{\log(\mu)} = 2 - \alpha \tag{I-16}$$

$$\lambda_3 = \lim_{\tau \to 0} \lim_{\pi \to 0} \frac{\log G(\tau, \mu)}{\log(\tau) \log(\pi)} = 0 \tag{I-17}$$

The above limits result in another two power relations, which are valid in the immediate vicinity of the critical points

$$G_\mu(\tau) \propto |\tau|^{-2-\alpha+2\beta} \tag{I-18}$$

$$G_\tau(\mu) \propto |\mu|^{2-\alpha} \tag{I-19}$$

Equations (I-18) – (I-19) represent the asymptotic behaviour of the function G(τ, μ) as $\tau \to 0$ or G(τ, μ) as $\mu \to 0$. At the microscopic level, we observe a singular behaviour at T = $T_c$ (Fig. 5a) and μ = $\mu_c$ (Fig. 5b – c).

For the bi-critical point, the exact relation β = 2 holds (compare with SI Table 1 and SI Figure 2), thus leaving us with the only free parameter (α) of the theory. Note a broad peak at the critical temperature (optimal growth occurs within a certain temperature range) and a sharp, more pointed peak for the critical chemical potential (a single value that corresponds to the optimum growth). In this representation, growth can be treated as a series of subsequent STs that take place in the cell wall (see Part II). It looks as though a cardinal quantity that is decisive for growth – in this particular approach – is the chemical potential. During the time of evolution of the extending cell wall, the chemical potential may form asymmetrical ratchet mechanism (ibid.). Oscillating biochemical reactions, which are common in cell dynamics, may be closely related to the emergence of the phenomenon of life itself (Martin et al., 2009). In this context, the relations for the chemical potential may also belong to the most fundamental dynamic constraints for the origin of life.

Complex systems such as the cell wall of a growing plant are solid and liquid-like at the same time. At the molecular level, they are both ordered and disordered (allowing for a transition between these states). There are obviously complex interactions between the model parameters and inputs from a lot of other kinds of inputs. However, the EoS describes the properties of the wall and unknown complex interactions in general way. Whether the structure – at a given time instant – is ordered or disordered, universal features can be identified using simple scaling laws.

**Discussion**

Erwin Schrödinger (1944) may have been the first to consider the thermodynamic constraints within which life evolves, thereby raising fundamental questions about organism evolution and development. Here, we attempted to present a description couched in terms of the EoS and methods borrowed from phase transitions theory. More precisely, this work seeks to model the relationship between plant growth and temperature (or pH) in terms of the EoS, thereby describing growth optima as functions of 'state' variables and critical exponents. Apparently, there is already a simple and compelling explanation for growth optima of plant growth from the biological point of view. Growth is mediated both directly and indirectly by enzymes and increases broadly in line with the frequently observed effect of temperature on enzyme activities (Berg et al., 2002). However, once temperatures are sufficient to denature proteins, enzyme activities rapidly decrease and plasma membranes and some other biological structures also experience damage (compare SI Figure 1b at a high temperature). In this context, the kink that is observed at T = $T_c$ in Fig. 4a may coincide with a temperature-driven ST, thus pointing to the critical temperature $T_c$, which is in agreement with the interpretation of the enhancement in the effective diffusion rates (Pietruszka, 2012; Pietruszka and Haduch-Sendecka, 2016). On the other hand, the plot in Fig. 4b resembles a "lambda" kind of STs in which the function F tends towards infinity as pH approaches the lambda point, which is similar to the heat capacity diffusive transition (Münster, 1969; Matlak and Pietruszka, 2001). Note, at the low-pH end, the plot does not approach zero, but tends to the finite value of one, thus naturally preferring a lower pH-value regime (below $pH_c$) for growth, as is predicted by acid growth hypothesis, while nonetheless allowing for limited (though diminished) growth at pH > $pH_c$.

*Sensing structural transitions*

A magnitude that infinitesimal changes result in symmetry change at phase transition always exist in continuous phase transitions (Landau and Lifshitz, 1980). Seemingly, such (order/disorder) symmetry changes (De Gennes, 1991) in biological systems may be connected with a mechanism (Rojas et al., 2011) by which chemically mediated deposition causes the turnover of cell wall cross-links, thereby facilitating mechanical deformation. It may also reflect the pectate structure and distortion that was suggested by Boyer (2009) (Fig. 5c, d) in the *Chara* cell walls by placing wall polymers in tension and making the load-bearing bonds susceptible to calcium loss and allowing polymer slippage that irreversibly deforms the wall. Proseus and Boyer (2007) suggested that the ladder-like structure would be susceptible to distortion and that the distortion would increase the distance between adjacent galacturonic residues (Fig. 5D in Boyer (2009)) and therefore the bonds may lengthen and thus weaken and decrease their affinity for $Ca^{2+}$. Dissociation may then occur, thus allowing an irreversible turgor-dependent expansion (see also Fig. 4 in Pietruszka (2013); this is also applicable for non-isochronous growth in pollen tubes). The direction of the maximal expansion rate is usually regulated by the direction of the net alignment among cellulose microfibrils, which overcomes the prevailing stress anisotropy (Baskin, 2005). However, the organization of cellulose microfibrils depends on the organization of cortical microtubules (Paredez et al., 2006), which is modified by many different factors, such as mechanical stimuli, light or hormones (Fishel and Dixit, 2013). These modifications are regulated *via* DELLA proteins (Locascio et al., 2013) and/or arabinogalactan proteins (Willats and Knox, 1996; Marzec et al., 2015b). The transient changes in the wall composition and the deformation of bonds may be a hallmark of symmetry change and a kind of ST (orchestrated instability) that is taking place in the cell wall.

The cell wall is being supplied with precursors of polysaccharides and some proteins also through exocytosis. This process is a form of active transport in which a cell transports molecules out of the cytoplasm to the extracellular matrix or intracellular space by expelling them in an energy-using process. Endocytosis is its counterpart. The loaded vesicles are transported by kinezine through microtubules into the plus end while by dinezine toward the minus ends, and ATP is utilized in both binding and unbinding processes of the motors heads. Besides microtubules, the actin cytoskeleton represents the most demanding energy-using process. In fact, in plant cells, both exocytosis and endocytosis are F-actin dependent processes related to cell wall assembly and maintenance; moreover, there is strong evidence that plant cell elongation itself also depends on actin cytoskeleton rearrangement, see a series of studies by Baluška et al. (1997, 2001 and 2002) and Wojtaszek et al. (2007). Indeed, these energy-consuming processes are apparently not included in the EoS.

A situation, which was already pointed out in the Introduction, in which the protons that are excreted into the wall compartment are directly responsible for wall-loosening processes through the hydrolysis of covalent bonds or the disruption of non-covalent bonds may be also a signature for the symmetry change and STs that occur in the wall compartment of the growing cell. Though the underlying microscopic mechanisms are not well recognised as yet (and as a result the 'order parameters' are difficult to identify), a pH-driven "lambda" ST may be attributed to the maximum activity of PM $H^+$-ATPase, while temperature-driven STs in the cell wall can be directly related to the maximum elevation of the effective diffusion rate ($k_2$ coefficient in Pietruszka (2012)) – we have considered primary (diffusive) growth throughout this article. As an aside, from the calculation of cross-correlations, we may draw further conclusions that are connected with the biochemical picture of the acid growth theory. The results presented in SI Figure 3 is a clear characteristic that temperature-induced growth and auxin-induced acidification (PM $H^+$-ATPase) growth correlate strikingly well (there is good experimental evidence that higher temperature induces auxin biosynthesis in some plants). However, the convolution of acidic pH growth and temperature growth is less shifted away from zero (lag) and is even more pronounced (see also recent outcomes in Pietruszka and Haduch-Sendecka (2016) for cell wall pH (proton efflux rate) and growth rate, which are directly co-regulated in growing shoot tissue, thus strongly supporting the empirical foundations of EoS). This strict quantitative result may contribute to the acid growth hypothesis that was developed by plant physiologists and make a contribution to resolving problems that occur in the long-lasting discussion. However, a problem still arises. It is unclear to what degree Eq. (I-5) applies universally as the published raw data that constitute the experimental basis are only for grass shoot elongation growth. The long-standing acid growth theory postulates that auxin triggers apoplast acidification, thereby activating cell wall-loosening enzymes that enable cell expansion in shoots. This model remains heavily debated in roots. However, recent results (Barbez et al., 2017) in investigating apoplastic pH at cellular resolution show the potential applicability of Eq. (I-5) for roots.

We now show that our findings may be embedded in the evolutionary context that is connected with the migration of plants away from the Equator (changes in latitudes) as the climate changed and their adaptation to the spatial distribution of pH in the soil as a substitute for high temperature. As an example that exhibits the capabilities of equation (I-5), let us consider the following. The dominant factors that control pH on the European scale are geology (crystalline bedrock) in combination with climate (temperature and precipitation) as was summarised in the GEMAS project account (Fabian et al., 2014). The GEMAS pH maps mainly reflect the natural site conditions on the European scale, whilst any anthropogenic impact is hardly detectable. The authors state that the results provide a unique set of homogeneous and spatially representative soil pH data for the European continent (note Fig. 5, ibid.). In this context, the EoS that is expressed by equation (I-5) may also gain an evolutionary dimension when considering the spatial distribution of pH data as affected by latitude (angular distance from the Equator) elevation on the globe, which is associated with climate reflection (ibid.). pH is strongly influenced by climate and substrate – the pH of the agricultural soils in southern Europe is one pH unit higher than that in northern Europe. See also Fig. 4 in Fabian et al. (2013), in which the pH ($CaCl_2$) of soil samples that are grouped by European climate zones are measured (descending from pH ≈ 7 (Mediterranean), pH ≈ 6 (temperate) through pH ≈ 5.25 (boreal) to pH ≈ 4.80 (sub-polar)). This fact may be reinterpreted in terms of pH – temperature duality that is considered in this work. When the migration of plants away from the Equator took place, lower pH values further to the North might have acted in a similar way as high temperatures at the tropic zones on the growth processes. This situation could have led to energetically favourable mechanisms (adaptation) such as the amplified ATP-powered $H^+$ extrusion into the expanding wall of plant cells and the secondary acidification of the environment (released protons decrease pH in the incubation medium), irrespective of the initial pH level, which is the main foundation for the acid growth hypothesis. Apparently, the evolutionary aspect of equation (I-5), which suggests a possible link with self-consistent adaptation processes ("The individual organism is not computed, or decoded; it is negotiated" (Walsh, 2010), as one of its first applications, seems difficult to underestimate, and the potentially predicative power of EoS looks impressive considering the modest input we made by deriving equation (I-5). In this perspective, the pH-response could potentially provide a plant with an adaptive advantage under unfavourable climate conditions. Together with quite recent discovery of phytochromes function as thermosensors (Jung et al., 2016), the EoS could be useful in the breeding of crops that are resilient to thermal stress and climate change.

*Perspectives*

The EoS for a biological system has not been reported until this study. To further verify the EoS and its implications, we note that based on the results that were obtained (calculated critical exponents), a kind of "universality hypothesis", which is known from the physics of phase transitions (Stanley, 1971), can be tested experimentally in order to extract "classes of universality" for plant species ($\Omega_S$), thereby substantiating the taxonomic divisions in the plant kingdom *via* quantitative measures (Table 1). This issue, which we may call "extended plant taxonomy", demands further experimentation and is beyond the scope of this theoretical work.

It seems the EoS may explain a number of phenomena and is powerful in its simplicity. It positions itself at the brink of biology and nano-materials. We believe it may be helpful in the identification of the processes of the self-assembly of wall polymers. In agricultural implementations, any departures from the optimum growth may simply be corrected by appropriately adjusting pH or T.

The finding of the EoS for plants adds an important dimension to the biophysical search for a better explanation of growth-related phenomena in a coherent way through its applicability to the visco-elastic (or visco-plastic) monotonically ascending and asymptotically saturated (Pietruszka, 2012) cell wall extension that is usually observed, as well as to pollen tube oscillatory growth (Part II). The underlying biochemical foundation, which is expressed in the form of acid growth hypothesis, could also help to understand the growth conditions that are ubiquitous in biological systems through coupling to EoS. In this respect, our method is accompanied by its partial validation – its application to the important biological question concerning the acid growth hypothesis. In the above context, it is not surprising that the growth of the cell wall in plants can be thought of as a time-driven series of STs (*via* the breaking of polymer bonds or other mechanisms that were mentioned earlier), although they are ultimately connected with physics and chemistry by the EoS (equations (I-5) or (I-6b)), which are solved at subsequent time instants by Nature. Since non-equilibrium stationary states are achieved after discrete intervals of time for single cells like pollen tubes, such a (chemical potential ratchet) mechanism can lead to a kind of 'leaps', macroscopically (of µm scale) observed growth oscillations (Geitmann and Cresti, 1998; Hepler et al., 2001; Zonia and Munnik, 2007). Similar mechanisms were observed in multicellular models for plant cell differentiation, such as rhizodermis (Marzec et al., 2013a) where growth of root hair tubes occurs on leaps and bounds way (Vazquez et al., 2014). This phenomenon may be treated as a final result of the subsequent STs that take place in the cell wall.

On the other hand, spontaneous bond polarisation/breaking may be synchronised through Pascal's principle by pressure P or pressure fluctuations δP (Pietruszka and Haduch-Sendecka, 2015a), thereby acting as a long-range (of the order of ξ) messenger at the phase change (the radius of coherence $\xi \to \infty$ at a Γ-interface (Pietruszka, 2013)). When the entire system is correlated and the stress-strain relations are fulfilled (ibid.), the simultaneous extension of the cell wall at the sub-apical region may correspond to pollen tube oscillation(s). Hence, the uncorrelated or weakly correlated extension would result in the bendings that are observed in the directional extension of the growing tube (ibid.). In multi-cell systems (tissues) that have a higher organisation (coleoptiles or hypocotyls in non-meristematic zones), the EoS for plants will manifest itself by the emergent action of an acidic incubation medium, or will be induced by endogenous auxin acidification, as is presumed in the acid growth theory. In this aspect, the EoS may serve as a new tool for the further investigation and verification of claims about the acid growth hypothesis. This should greatly facilitate the analysis of auxin-mediated cell elongation as well as provide insight into the environmental regulation of auxin metabolism. Moreover, using the EoS it will be possible to answer the question about role of crosstalk between auxin and other hormones, such as strigolactones (Marzec and Muszynska, 2015), in growth of plant cells. It also seemingly delivers a narrative that provides a biophysical context for understanding the evolution of the apoplast, thereby uncovering hidden treasures in the as yet unscripted biophysical control systems in plants.

**Summary**

This part takes a statistical physics approach to plant growth, looking for simplifying universal features that may explain observations. We have fitted a simple algebraic function (a beta Euler distribution) to some growth rate data, capturing variations with respect to temperature and pH, which show a surprising conjugate nature.

Cross-disciplinary research at the interface between the physical and life sciences was accomplished in this part. We first considered the Euler beta function enigma, which is a duality of acidic pH (or auxin-induced acidification) and temperature, in order to obtain the EoS for living plants. We started from the striking similarity (mirror symmetry) between the elongation growth of grass shoots that were incubated in different pH environments or auxin-induced elongation in water (or a neutral buffer) with the respective growth at different temperatures. We based these on the hypothesis that the action of temperature on the elongation growth is effectively equivalent for the relative growth increments such as those that are caused by the change of pH or endogenous auxin-induced acidification in the incubation medium. In order to resolve this ambiguity, we first used the beta function for both dependences, which were normalised prior to the comparison, in order to obtain the beta function shape parameters α and β. It turned out that even without referring to biochemical underpinning, we might have concluded that the acidic conditions of the incubation medium or auxin-induced acidification and environmental temperature can act interchangeably, at least when they are considered at (mechanistic) phenomenological level, i.e. from the effectiveness of the plant growth point of view. The numerically verified high accuracy of these complementary representations allowed us, among others, to treat this dual approach as a new apparatus for predicting the outcomes in the swapping growth conditions, which is especially useful in changing climate surroundings. We presumed that by applying a beta distribution and continuously changing its character with the values of the shape parameters, our findings might be related to the evolutionary context that is connected with the migration of plants away from the Equator as the climate changed and their adaptation to spatial distribution of pH in the soil as a substitute for high temperature. The EoS may be also helpful in delivering analytic solutions (the respective (α, β, $\Omega_S$) – triads) for the assisted migration of plant species (Vitt et al., 2010) that are at risk of extinction in the face of rapid climate change. In this application, the EoS tool can constitute a complementary method for genetic modifications in assisted migration processes, by choosing species for such an introduction from those having a similar (to within experimental error) triad.

In physics, it is believed (Fisher, 1966; Griffiths, 1970) that the critical exponents are "universal", i.e. independent of the details of the Hamiltonian (energy function) that describes a system. The implications of this, however, are far reaching. One could take a realistic and complicated Hamiltonian, 'strip' it to a highly idealised Hamiltonian, and still obtain precisely the same critical exponents. For instance, on these grounds, it is believed that carbon dioxide, xenon and the three-dimensional Ising model should all have the same critical exponents (Baxter, 1989). This appears to be the case (Hocken and Moldover, 1976). In this context, since growth phenomenology and the STs approach have converged to a form for a EoS for plants, the potential role of critical exponents to discriminate the different modes of wall extension (Breidwood et al., 2014 for review) of growing cells or tissues seems promising in opening new avenues of research.

## PART II

## Equation of Evolution for Plants

**Introduction**

In this part we pose a question whether the elongation growth of plants (represented by expanding volume V) can be entirely predicted by pH, T and P?

The modulation of mechanical properties during expansive growth is a 'hot topic' for plant cell growth community, e.g., recently Bidhendi and Geitmann (2015), Boudon et al. (2015), but also in Rojas et al. (2011) and in Barbacci et al. (2013). Plant cell expansion, in turn, is controlled by the balance between intracellular turgor pressure, cell wall synthesis and stress relaxation. Several mainstream explanations of the mechanical aspects of cell growth have been considered in publications, beginning with the cell wall model (Holdaway-Clarke and Hepler, 2003) through the instability model (Wei and Lintilhac, 2007) to the hydrodynamic model (Zonia and Munnik, 2007). In this part, we advocate that these apparently conflicting visions of cell wall extension and growth may be consolidated in a unifying chemical potential based explanation, which hopefully converge to a coherent picture of this somewhat elusive object.

The last few years have witnessed significant progress in uncovering the molecular basis underlying plant cell elongation and expansion. In many cases, the first step in cell growth is related to a restriction of the symplasmic communication between neighbouring cells, which allows the specialization programme that is specific for the individual cell to begin (Marzec and Kurczynska, 2014). This isolation enables cells from multicellular organisms to be considered as individual cells. Afterwards, different signalling and developmental mechanisms are run to induce either the expansion of the entire cell or localised elongation. The role of different protein classes have been described based on the analysis of the growth of pollen or root hair tubes, which are the models used in investigations of tip-growth (Higashiyama et al., 2015; Marzec et al. 2015a). Among them, the function of expansins, the proteins that are involved in plant cell wall loosening, which is related to cell wall acidification, was confirmed in both monocot and dicot species. Analysis of different mutants in the genes encoding expansins confirmed their role based on the analysis of plant phenotypes (Cosgrove, 2015). Nevertheless, a model that allows cell status of mutant and wild-type plants to be compared might prove useful for further investigations.

According to the acid growth hypothesis, auxin plays the pivotal role in growth of plant cell (Hager, 2003; Lüthen, 2015 for review). However, recent reports have indicated that a new class of plant hormones, strigolactones (SLs), cooperate with auxin in many developmental processes such as shoot and root branching under normal and stress conditions (Marzec et al., 2013b). Additionally, new functions of SLs have recently been proposed, which indicate that those hormones may play a key role in the regulation of plant growth and development (Marzec and Muszynska, 2015). Until now, most data has indicated that SLs influence auxin transport *via* the distribution of PIN proteins (Shinohara et al., 2013), but it has also been proposed that SLs influence secondary growth in plants (Agusti et al., 2011) and cell elongation (Hu et al., 2010; Koltai et al., 2010). Mathematical formulas that allow cell growth under different treatment to be measured and described may shed new light on the hormonal regulation of plant cell elongation.

From the physical point of view, the basic ingredients of an extending plant cell are the cell wall, plasma membrane and cytoplasm. Together, they form a system that can be described as a two-compartment structure with the plasma membrane being treated as the interface between the cytoplasm and the wall. On the other hand, the intense quantities can be attributed to the system of volume V, namely pressure (P), and the chemical potential (µ), the latter of which is directly related to a measurable quantity – pH. The problem of the volumetric growth of plant cells, which lies at the intersection of physics, biochemistry and composite materials of a nanometric scale, is inherently difficult to capture in a single theoretical frame (Braidwood et al., 2014). Therefore, the goal was to deconstruct all of the above-mentioned approaches in order to assemble a model that is based on physical principles that are also able to admit other perspectives. We believe that the proposed above EoS has the ability to do this, although its relevance to many of the growth details may be beyond the scope of this conjecture.

*Chemical wall loosening approach*

It is believed that the cell wall possesses active mechanisms for self regulation (Hepler and Winship, 2010). The cell wall is formed through the activity of the cytoplasm and maintains a close relationship with it. Communication occurs between the cell wall and the cytoplasm and the plasma membrane occupies a pivotal position in transmitting particles and information. In this model, when applied to pollen tube growth, cell growth is ultimately dependent on the biochemical modification of the wall at the apical part where exocytosis is believed to occur (Holdaway-Clarke and Hepler, 2003). Cell wall loosening then allows turgor-induced stretching, thereby pulling the flow of water into the cell (down its potential gradient). During oscillatory growth, enzymes that interfere with wall loosening are believed to be periodically activated or inhibited, which is supposed to drive oscillations in growth and the growth rate (as commented by Zonia, 2010). In this approach, the passive role of turgor pressure is stressed.

*Loss of stability model*

A new look at the physics of cell wall behaviour during plant cell growth, which focuses on the purely mechanical aspects of extension growth such as wall stresses, strains and cell geometry, was investigated within the framework of a model based on the Eulerian concept of instability (Wei and Lintilhac, 2003; 2007). It was demonstrated that loss of stability (LOS) and consequent growth are the inevitable result of the increasing internal pressure in a cylindrical cell (like the internodal cylindrical cells of intact *Chara corralina* plants) once a critical level of pressure-induced stress ($P_c$) is reached. The predictions of this approach were confirmed by direct measurements (ibid.). This model (LOS) apparently challenged the viscoelastic or creep-based models (Ray et al., 1972; Cosgrove, 1993; Schopfer, 2006), which were designated as chemical wall loosening and was questioned by Schopfer (2008), who asserted that stress relaxation can be attributed to chemorheological changes in the load-bearing polymer network that permit the plastic deformation of wall dimensions. In his words, "growth of turgid cells is initiated and maintained by chemical modifications of the cell wall (wall loosening) followed by mechanical stress relaxation generating a driving force for osmotic water uptake". In response to Schopfer's letter, Wei and Lintilhac (2008) claimed that LOS did not challenge the principles of osmotic water relations or biochemically mediated wall loosening and emphasised that plant cells are both osmometers and pressure vessels, and therefore, both facts should be accommodated in any meaningful model. However, the problem of the water potential difference $\Delta\Psi = \Psi_o - \Psi_i$ of a growing cell and a "mysterious pump" if pressures must rise to drive LOS behaviour still remained. Regardless of whether the new wall synthesis is homogeneous or patchy (the cell wall can always be conceptually shrunk to a patch), it will affect cell behaviour by modifying the turgor pressure P in short time scale oscillations or pseudo-chaotic fluctuations $\delta P$ (Pietruszka and Haduch-Sendecka, 2015a), which are specifically enhanced by flow of the osmotic pressure through the entire cell volume (beyond the apex as well).

*Hydrodynamic model*

Recent works have revealed an important role for osmotic pressure and the hydraulic features of the system in driving cell shape restructuring and growth (Proseus et al., 2000; Proseus and Boyer, 2006; Zonia et al., 2006, 2010; Haduch-Sendecka et al., 2014). This concept was established by works that revealed an important role for hydrodynamics in pollen tube growth. The main message that hydrodynamics have the potential to integrate and synchronise the function of the broader signalling network (in pollen tubes) was expressed by Zonia (2010). To be accurate, Pascal's principle states that pressure exerted anywhere in a confined incompressible fluid (cell sap) is transmitted equally in all directions throughout the fluid in such a way that the pressure variations (initial differences) remain the same. Indeed, according to Pascal's principle, turgor pressure is a fast (propagation equal to the velocity of sound in the fluid – soap – medium) and unperturbed (isotropic) messenger in the time-evolving cell. Moreover, hypotonic or hypertonic treatment of the periodically elongating cells of pollen tubes led to the recognition that the growth (growth rate) oscillation frequency was "doubled" or "halved" with respect to the unperturbed (isotonic) mode (ibid.). The oscillation frequency in such a system has been determined (Pietruszka, 2013) as obeying the following relation: $\omega \sim \sqrt{P}$, which means that the oscillation growth angular frequency is directly related to the pressure.

From these brief accounts, the question arises – What drives plant cell growth – cell wall loosening or osmotic pressure? Even though there is a consent about the role of cell wall properties and turgor pressure in growth, a discrepancy about what drives the initial episode of cell extension exists (Zonia and Munnik, 2007). One model proposes that cell wall loosening occurs first and permits the expansion of the cell due to decreased wall rigidity. There is considerable evidence to support this view (Cosgrove, 2005b). However, cytomechanical studies have revealed that the viscoelastic properties of the pollen tube cell wall in the apical growth zone do not change considerably (Geitmann and Parre, 2004) during growth oscillations. Hence, it was suggested that the cycles of cell wall loosening cannot be the driving force of oscillatory growth (Zonia and Munnik, 2007). Moreover, it was shown that increased pressure induces the expansion of the cell wall and drives growth in *Chara* cells (Proseus et al., 2000; Proseus and Boyer, 2006). An intermediate LOS model has been put forward (Wei and Lintilhac, 2003; Wei et al., 2006) that proposes a gradual increase of turgor pressure to a critical point $P_c$ at which the loss of stability initiates wall extension and growth. Without going further into the details of the ongoing discussion (Zonia and Munnik, 2007; Winship et al., 2010; Winship et al., 2011) and bring together these different opinions, we propose a chemical potential based scenario. In what follows, we have intentionally omitted irrelevant details in order to make the main steps clear while introducing it and to build a story with logical a flow.

**Results**

*Mystery of the Lockhart coefficient explained?*

If one imagines that the growth rate, as a function of pH, has a (Euler) beta function form for any fixed T and a similar dependence as a function of T for fixed pH (see SI Table 1 and SI Figure 1), one can conclude that the volumetric growth rate is proportional to a product of the beta functions of T and pH (and to the 'effective' turgor pressure P – Y). This assertion can be a starting point for 'derivation' (in first approximation, where we neglect the nested pH($\mu_{H+}$(T)) dependence) of the equation beneath. Simply speaking, the above statement for the relative growth rate can be expressed in an analogous way to the Lockhart (1965) equation, as an Ansatz ("educated guess")

$$\underbrace{\frac{1}{V(t)}\frac{dV(t)}{dt}}_{\text{volume relative growth rate}} = \underbrace{\underbrace{[c_{\text{pH}} \times \text{pH(t)}^{\beta-1}(1-\text{pH(t)})^{\alpha-1}]_T}_{\text{pH-dependent\_beta\_function}} \underbrace{[c_T \times T^{\alpha-1}(1-T)^{\beta-1}]_{\text{pH}}}_{T\text{-dependent\_beta\_function}} \times \Phi_0}_{\text{Lockhart constant } \Phi} \times \underbrace{(P-Y)}_{\text{effective pressure}} \qquad \text{(II-1)}$$

at a constant turgor pressure P and yield threshold Y (both in [MPa]; we assume concomitant water uptake); $c_T$ and $c_{pH}$ are beta function scaling constants. The method that was used to normalise pH and T to [0, 1] interval was described in detail in Part I – in fact it consisted simply of division by 14 for pH scale, and by the maximum growth temperature (we postulated $T_{max}$ = 45 $^{o}$C) for T scale, respectively. V = V(t) is the increasing volume of a growing cell over the course of time. Here, "pH" is treated, as before, as a non-separable variable name and a constant $\Phi_0$ was introduced for dimensionality reasons – $[\Phi_0]$ = $10^{-6}$ MPa$^{-1}$ × s$^{-1}$, like in the Lockhart equation. Integration of Eq. (II-1) yields the solution for the enlarging volume in the form

$$\int_0^\tau \frac{dV(t)}{V(t)} = [c_T]_{pH} \times [T^{\alpha-1}(1-T)^{\beta-1}]_{pH} \times \Phi_0 \times (P-Y) \times [c_{pH}]_T \int_0^\tau [\text{pH}(t)^{\beta-1}(1-\text{pH}(t))^{\alpha-1}]_T dt \qquad \text{(II-2)}$$

where 0 and τ denotes the initial and final time. In order to determine the unknown constants we can part the whole period τ into four subintervals, approximate by polynomial splines (Fig. 6)

$$\int_0^\tau [\ldots] \equiv \int_0^{\tau_1} [\ldots] + \int_{\tau_1}^{\tau_2} [\ldots] + \int_{\tau_3}^{\tau_4} [\ldots] + \int_{\tau_4}^{\tau} [\ldots] \qquad \text{(II-3)}$$

and solve the system of equations for α, β, $c_T$ and $c_{pH}$ (see beneath).

Note that, unlike the Lockhart equation, "acidification" present in Eq. (II-2) – represented by the value of pH – prevents a plant cell from unlimited exponential growth (burst). Equation (II-2) properly delivers not only the initial phase of slow, and next of rapid growth (termed "inflation phase" by the author), but also deceleration and saturation phases (Fig. 7). Equation (II-1) holds for dynamic pH (in Fig. 8 changes in the volume and saturation effect depend on simulation parameters).

By recalling the canonical form of the Lockhart equation, with the extensibility Φ controlling the rate of growth (proportional to the excess of turgor pressure P – Y)

$$\frac{1}{V(t)} \frac{dV(t)}{dt} = \Phi(P - Y) \qquad \text{(II-4)}$$

and comparing Eqs (II-1) and (II-4), we get the estimate for the Lockhart 'constant':

$$\boxed{\Phi = \Phi(\text{pH}, T) = [c_{pH} \cdot \text{pH}^{\beta-1}(1-\text{pH})^{\alpha-1}]_T [c_T \cdot T^{\alpha-1}(1-T)^{\beta-1}]_{pH} \Phi_0} \qquad \text{(II-5)}$$

where the exponents α and β are numbers that are independent of pH (or μ) and T, whereas pH = pH(t). Hence, the Lockhart coefficient Φ, responsible for visco-plastic extensibility, is an analytic function of pH and T on the physically accessible interval [0, 1] for each variable (Part I). However, the augmented form of the Lockhart equation, which is expressed by Eq. (II-1), provides the possibility to not only describe monotonic cell expansion properly (Figs 7-8), but also the oscillatory mode of the extension of the pollen tubes. In addition, in the case of oscillatory turgor pressure (or just turgor pressure fluctuations like in Pietruszka and Haduch-Sendecka, 2015a), the additional (visco-elastic) term 1/ε × dP(t)/dt (Ortega, 1985) can be introduced (added) in equation (II-1) on the right side, for completeness. Also, the remarks concerning possible variability of Y during growth (Pietruszka, 2012) are relevant in the higher order of approximation. Determination of α and β shape coefficients for the actual empirical data is presented beneath.

### *Determination of α and β shape coefficients for experimental data*

Potential applications of our formalism lie in α and β parameters. For usual experimental conditions (T = const.), by taking the natural logarithm of Eq. (II.1), we get

$$\log\left(\frac{1}{V}\frac{dV}{dt}\right) = \log c_T + (\alpha-1)\log T + (\beta-1)\log(1-T) + \log c_{pH} + (\beta-1)\log pH + (\alpha-1)\log(1-pH) + \log \Phi_0 + \log(P-Y)$$

(II-6)

Multiplying both sides by d$t$ and integrating over the subsequent subintervals, we receive the set of linear equations to determine α, β and $c_T$, $c_{pH}$. In a matrix form we have

$$\mathbf{Ax} = \mathbf{B} \qquad (\text{II-7})$$

where, explicitly,

$$\mathbf{A} = \begin{pmatrix} \int_0^{\tau_1}[\log T + \log(1-pH(t))]dt & \int_0^{\tau_1}[\log pH(t) + \log(1-T)]dt & \tau_1 & \tau_1 \\ \int_{\tau_1}^{\tau_2}[\log T + \log(1-pH(t))]dt & \int_{\tau_1}^{\tau_2}[\log pH(t) + \log(1-T)]dt & \tau_2-\tau_1 & \tau_2-\tau_1 \\ \int_{\tau_2}^{\tau_3}[\log T + \log(1-pH(t))]dt & \int_{\tau_2}^{\tau_3}[\log pH(t) + \log(1-T)]dt & \tau_3-\tau_2 & \tau_3-\tau_2 \\ \int_{\tau_3}^{\tau}[\log T + \log(1-pH(t))]dt & \int_{\tau_3}^{\tau}[\log pH(t) + \log(1-T)]dt & \tau-\tau_3 & \tau-\tau_3 \end{pmatrix}$$

$$\mathbf{x} = \begin{pmatrix} \alpha \\ \beta \\ \log c_T \\ \log c_{pH} \end{pmatrix}$$

and

$$\mathbf{B} = \begin{pmatrix} \int_0^{\tau_1} \log\left(\frac{1}{V(t)}\frac{dV(t)}{dt}\right)dt + \int_0^{\tau_1}[\log T + \log pH(t) + \log(1-T) + \log(1-pH(t)) - \log\Phi_0 - \log(P-Y)]dt \\ \int_{\tau_1}^{\tau_2} \log\left(\frac{1}{V(t)}\frac{dV(t)}{dt}\right)dt + \int_{\tau_1}^{\tau_2}[\log T + \log pH(t) + \log(1-T) + \log(1-pH(t)) - \log\Phi_0 - \log(P-Y)]dt \\ \int_{\tau_2}^{\tau_3} \log\left(\frac{1}{V(t)}\frac{dV(t)}{dt}\right)dt + \int_{\tau_2}^{\tau_3}[\log T + \log pH(t) + \log(1-T) + \log(1-pH(t)) - \log\Phi_0 - \log(P-Y)]dt \\ \int_{\tau_3}^{\tau} \log\left(\frac{1}{V(t)}\frac{dV(t)}{dt}\right)dt + \int_{\tau_3}^{\tau}[\log T + \log pH(t) + \log(1-T) + \log(1-pH(t)) - \log\Phi_0 - \log(P-Y)]dt \end{pmatrix}$$

where pH = pH(t) scaled is known from experiment. In general, the system temperature $T$ or turgor pressure $P$ may also change in time. The vector of solutions (**x**) can be obtained with the use of any linear equation solver algorithm, e.g., by using SageMath, which is a free open-source mathematics software system licensed under the GPL, or many others. However, by admitting $\alpha + \beta \geq 2$ (Part I) we can pose a constraint for the shape variables at the isotherm of the optimum growth. Further, the product $[c_{pH}]_T \times [c_T]_{pH}$ can be cautiously replaced by one constant, thus leaving us with the set of two linear equations, dimension(A) = 2 x 2. However, calculation performed for the subsequent (similar to the moving average method) more dense subintervals (e.g., every 15 min., as in our case) – in the whole considered range – should give a better estimation for <$\alpha$> and <$\beta$> and the vales of uncertainties (errors) for $\alpha$ and $\beta$. Here we used a brute force method for qualitative analysis of the problem. The quantitative statistical analysis is quite involved and would be presented in a following/separate paper. Then we would explain in detail how the model can be calibrated using specific experimental data. Also, biological meaning of the power law exponents that is not discussed in detail here, should appear for more abundant sets of data.

*Monotonic cell elongation*

In order to check our Ansatz, which is expressed by Eq. (II-1), we examined the pH decrease that were induced by auxin (IAA) and fusicoccin (FC) in Lüthen et al. (1990), which is presented in Fig. 6 therein. There, the additions of IAA and FC resulted in rapid drops of pH. The authors asserted that these lowered pH values are responsible for the fast increase in the growth rate. We digitised the pH data that was presented by Lüthen et al. (1990) and inserted them separately (point by point) into Eq. (II-2). As a result, we received two fragments of ideal growth curves (possessing exceptionally high determination coefficients!), which are presented in Fig. 9. (Note that to determine the cell volume evolution in time, we only needed the values of the following dynamical variables: pH, temperature and turgor pressure; the initial volume $V_0$ was needed for scaling; T = const. and P ≈ const., though it may also change in time). Interestingly, the simulation data for the $\alpha$ and $\beta$

parameters – representing the properties of the wall – were adapted from Part I, combining the EoS with the evolution equation (II-2). These data, for which Eq. (II-2) is not divergent (unlike to the Lockhart equation), closed the 'system of equations' for growth (here: for maize coleoptiles' elongating cells). By the 'system' we mean (a) the EoS, which describes the properties of the cell wall, irrespective of the extension mechanism, and (b) the equation of time evolution, which is specified here by Eq. (II-2), which describes the dynamics of the growing system at temperature T of the environment (see also SI Figure 6).

In this paragraph we have demonstrated that equation (II-2) correctly describes monotonic growth in the context of the "acid growth hypothesis" (note that pH is included explicitly in this equation; see also Figs 8 – 9 and Eq. (9) in Pietruszka and Haduch-Sendecka (2016) and conclusions therein) unambiguously confirming its long discussed applicability not only for fusicoccin, but also for auxin (Hager, 2003). The extremely high accuracy of theoretical fits (solid lines) with experimentally originated (Fig. 6 in Lüthen et al., 1990) data, presented in Fig. 9, strongly support the presented approach. However, the question about possible applicability of Eq. (II-2) to oscillatory growth still remains.

*Periodic cell elongation*

It is known that the group of oscillating variables during growth includes at least ion fluxes (see Pietruszka and Haduch-Sendecka (2015b) and Pietruszka et al. (2017) for oscillation time series, frequencies and intensities) and internal free concentrations of $Ca^{2+}$, $Cl^-$, $H^+$ and $K^+$ ions (Holdaway-Clarke et al., 1997; Holdaway-Clarke and Hepler, 2003), the cytoskeleton (Geitmann, 2010), membrane trafficking (Feijo, 1999; Feijo et al., 2001; Zonia and Munnik, 2008) and cell wall synthesis (see Portes et al., 2015 and Hemelryck et al., 2017 for review). However, despite the progress made in this domain, a central core-controlling mechanism is missing. Moreover, how it produces a macroscopic outcome in the form of apical growth is unknown. A possible link between this oscillatory component at the molecular level and structurally and temporally organised apical (periodic) growth is proposed beneath.

The pH value is a logarithmic measure of $H^+$-activity (the tendency of a solution to take $H^+$) in aqueous solutions. However, pH can be directly expressed in terms of the chemical potential ($\mu_{H^+}$) of protons through the relation

$$\text{pH} = \log_{10}\left[\exp\left(-\frac{\mu_{H^+} - \mu_{H^+}^0}{RT}\right)\right] \tag{II-8}$$

where R denotes the gas constant, T stands for the absolute temperature and $\mu_{H^+}^0$ is the reference potential. In this context, pH is not a typical generalised coordinate, although the microscopic state of the system can be expressed through it in a collective manner. Just as temperature determines the flow of energy, the chemical potential determines the diffusion of particles (Baierlein, 2001).

After inserting Eq. (II-8) into Eq. (II-2), we can calculate the influence of the chemical potential oscillations of $H^+$ ions, $\mu_{H+} = \mu_{H+}(t)$, on the volumetric expansion. A naive numerical simulation reveals that the volumetric extension represents a typical growth (sigmoid) curve (like the one presented in Fig. 7b). However, in this case, when superimposed onto the chemical potential the simulated oscillations produce growth rate oscillations in a straightforward manner (SI Figure 4). This type of oscillations was observed in the fast expansion of the pollen tubes (Plyushch et al., 1995; Hepler et al., 2001; Feijo et al., 2001; see also Pietruszka et al., 2017 for pH oscillations during growth), and the source of this periodic motion has been the subject of a long lasting debate among plant biologists. Portes et al. (2015) noted that "Since the ultradian rhythms present in pollen tubes are not temporally coincident with protein synthesis and degradation, a simpler mechanism might be triggering the overall oscillation". Further, the authors claimed that "the strongest candidates seem to be the membrane trafficking machinery and ion dynamics, since they are the only candidate mechanisms that may oscillate independently of tube extension (Parton et al. 2003)". Indeed, in what follows a chemical potential-induced (ultimately connected with ion dynamics since it measures the tendency of particles to diffuse) recurrent model, which triggers the overall oscillation, is proposed to explain this oscillatory mode of a single cell extension.

**Discussion**

Indirect measurements of the chemical potential ($\mu_{H+}$) by means of pH can be compared to our previous measurements of the electromotive force (EMF) in order to detect phase transitions, which were localised by the 'kinks' in the chemical potential, in condensed matter physics (Matlak and Pietruszka, 2001; Matlak et al. 2001, van der Marel 2004, Rówiński and Pławecki, 2016 patent). Such 'kinks' in the chemical potential always localised the occurrences of different types of phase transitions (where the change of state of the macroscopic system occurs), such as ferromagnetic, anti-ferromagnetic, superconducting, re-entrant or structural phase transitions (Matlak et al., 2000). Based on these observations and considerations on the EoS for plants, we can combine these outcomes together in a single coherent account.

*Chemical potential-induced ratchet-like growth in plants*

Launching our reasoning from the EoS and putting the differing approaches that were outlined in the Introduction together, we can construct a plant cell expansion recurring model. The latter, though simplistic for obvious reasons, not only describes one cell growth cycle in physical (mechanistic) terms, but also demonstrates how subsequent growth cycles can arise and – even – how the whole process may eventually decay (compare with Rounds et al., 2010).

The hypothetical steps in the transport of cations (such as $H^+$) against its chemical gradient by an electrogenic pump and its consequences are presented in Fig. 10 and SI Figure 5. These can be described as follows.

A. The initial state of a growing cell (we assume the chemical potential $\mu = \mu_0$ for reference only).

B. The outward, active proton transport across the plasma membrane interface creates gradients of pH or equivalently gradients of the chemical potential $\mu = \mu_{H+}$, which drive the transport of ions and uncharged solutes. The plasma membrane (marked by a vertical line) depicts the boundary between the cell wall and the cytoplasm.

C. $H^+$-ATPase activity regulates cytoplasmatic pH (and controls cell turgor pressure). At the microscopic level, we can observe a singular behaviour at a certain critical value of the chemical potential $\mu = \mu_c$ (see also Fig. 5c). This chemical potential activity can be interpreted as a signature for a ST that is occurring in the cell wall – during a ST of a given medium, certain properties of the medium change, often discontinuously, as a result of the change in some condition, here: $\mu$.

D. The increased involvement of $H^+$-ions in cell wall building and exocytosis (Holdaway-Clarke and Hepler, 2003) by $\delta V$ raises the turgor pressure inside the cell by $\delta P$, thus producing the force F, which may cause cell wall loosening (loss of stability, Wei and Lintilhac, 2007) and elongation by $\delta l$. The physical quantity that can be responsible for signal transduction here is pressure (e.g., Zonia, 2010), which is, however, a scalar quantity.

E. The increased pressure at the apical plasma membrane by $\delta P$ induces the stress relaxation of the cell wall, relaxation-driven cell expansion by $\delta V$ and triggers the start of the next growth cycle with a slightly modified turgor pressure $\sim P - \delta P$ (see the linearly descending pressure plotted in Fig. 1 in Benkert et al., 1997). The system restores the initial configuration and allows a new growth cycle to begin (explicit in pollen tube oscillations).

As a result, the problem of causality, which was mentioned in the Introduction, is resolved in a self-consistent manner thus potentially leading to a paradigm change expressed by a selfconsistently evolving recurring model. Obviously, the latter should be further examined by numerical calculations (self-consistent solutions at subsequent time intervals). Possible mechanisms of the participation of pH changes on photosynthesis (Grams et al., 2009; Sukhov et al., 2014; Sherstneva et al., 2015; Sukhov et al., 2016) can also be included in the form of a phase-locked loop, thus closing the system of equations. The existence of oscillating behaviours in the investigated system, however, implies the existence of a control by feed forward or feedback loops, which are inserted into a self-regulating mechanism (Portes et al., 2015). The latter statement can be translated in mathematical (physical) terms into a self-consistent evolution equation that corresponds with growth.

We will briefly discuss the major novelty of the recurrent model as compared to the earlier work in the context of the previous models. The three main approaches that were outlined in the Introduction emphasised a single attribute that is responsible for expansion (or periodic) growth. However, it seems that none of these factors alone (such as oscillatory wall synthesis, excessive turgor pressure or turgor pressure oscillations) can explain all of the observable modes of cell expansion and growth. Moreover, because of the intricate dependence between the processes that are involved in tip growth, causal relations cannot be easily inferred solely from phase relations (Pietruszka and Haduch-Sendecka, 2015b), as simultaneous information about the role of participating ions and the actual growth rate amplitude is required (ibid).

Furthermore, intracellular ion dynamics has been shown to play an essential role in pollen tube growth (Portes et al., 2015), while calcium, protons and chloride are also considered to be of major importance. Because the $[Ca^{2+}]_i$ pool at the apex oscillates (ibid.), it has also been considered as supporting growth (Messerli and Robinson, 1997). Additionally, it has been experimentally verified that pollen tubes cease growing when the $[Ca^{2+}]_i$ gradient is dissipated (Pierson et al., 1994). An interesting fact regarding chloride dynamics is that in contrast to $Ca^{2+}$ and $H^+$, the ionic fluxes occur in the reverse direction at the apex thereby elevating the chemical potential gradients even further.

Protons have received much attention as they play a crucial role in pollen tube growth. Since water, which ionizes spontaneously, is the major component of living cells, $H^+$ ions are known to be involved in enzymatic activity, endo/exocytosis and many others (Portes et al., 2015 for review). Therefore, pH inside cells has to be regulated in harmony with the growth phases (Messerli and Robinson, 1998). It was found that the pollen tubes showed an acidic domain located close to the membrane apex – the latter was present only during growth. In addition, the localisation of pollen-specific $H^+$-ATPase provides evidence indicating correlations with proton extracellular fluxes and the acidic domain at the apex (Certal et al., 2008). It looks as though proton dynamics is involved in the mechanisms for maintaining the polarity that ensures pollen tube growth (Portes et al., 2015). Moreover, it has been established (Agudelo et al. 2016) that the pollen tube response correlated with the conductivity of the growth medium under different AC frequencies—consistent with the notion that the effect of the field on pollen tube growth may be mediated *via* its effect on the motion of ions. This is, however, in accord with the presented chemical potential (ratchet) scenario and the representative time series of oscillatory traces of incoming cations ($H^+$ and $Ca^{2+}$) or outgoing anions ($Cl^-$) at the growing tip. Hence, from point of view of the recurrent model, the equilibration of the chemical potential for all of these species should also be the major mechanism for the cessation of growth.

Ion homeostasis and signalling are crucial to regulate pollen tube growth and morphogenesis, and affect upstream membrane transporters and downstream targets (Michard et al., 2016). Pollen tube growth is strictly dependent on ion dynamics: Ion fluxes and cytosolic gradients of concentration have been mechanistically associated with the action of specific transporters, especially for protons (ibid.). In the recurrent model put forward in this part, though extremely simplistic, chemical potential dynamics (Pietruszka et al., 2017) may serve as a "still missing" (Portes et al., 2015) central core-controlling mechanism that is able to produce a macroscopic outcome, i.e. structurally and temporally organised apical growth.

**Summary**

This part aims to develop a unifying model to simulate plant cell expansion and growth. In the chemical potential ($\mu$) representation, growth can be treated as a dynamic series of STs that can take place in the primary plant cell walls (Part I) in subsequent time intervals. We suppose that such changes would proceed, e.g., from an ordered to disordered state, from a stressed to relaxed state, from a "covalent bonds state" to "disrupted covalent bonds state", or even through more subtle pathways in the wall polymer network, which constitute an intriguing research task themselves. Furthermore, the singular causality, whereby only biochemical wall loosening can lead to stress relaxation and turgor loss, is replaced by a self-consistent recurring model, in which all of the prevailing variables constitute the macroscopic evolution of the plant cell. A common denominator appears to emerge that ties all of the growth factors together, namely the chemical potential.

At the beginning of this part we asked, if the elongation growth (V) of plants could be calculated from pH, T and P? It turned out that the answer is "yes", provided that it is delivered *a posteriori*, i.e., pH growth-data is already known. This reasoning, however, can be reversed: by knowing pH, T and P we can predict the cell volume V expansion in time using a model equation II-2.

**Conclusions**

By resolving the duality of low pH or auxin action (producing acidic pH) against temperature, not only have we introduced EoS for the realm of plants, but also, by considering wall extension growth as a dynamic cascade of chemical potential driven STs, identified critical exponents for this phenomenon, which exhibit a singular behaviour at critical temperature and critical pH in the form of power laws. A common unifying principle, which can be applied either for condensed matter – studied in physics, or living matter – being a subject of biology, is proposed for all chemical potential driven transitions.

Furthermore, universal (and exact) scaling relations were introduced that hold at the bi-critical point, which is in agreement with the experiment. The EoS (and evolution equation), which is strongly predicative, can either be helpful for resolving food resource problems on Earth – facing climate changes (exploit it in agriculture), or could be used as an auxiliary tool (calculator) for optimising fresh food production in the manned exploration of space.

**Table caption**

**Table 1** Preliminary data for α and β for maize (*Zea mays* L.) and wheat (*Triticum vulgare* Vill.) belong to the same family *Poaceae* but to the different subfamily: *Panicoideae* and *Pooideae*, respectively; the calculated values for α and β are similar. Common bean (*Phaseolus vulgaris* L.) and cucurbita (*Cucurbita pepo* L.) belong to the same clade *Eudicots* but to the other order: *Faboles* and *Cucurbitales*, respectively; the calculated values for α and β vastly differ. Calculation based on Fig. 1 in Lewicka and Pietruszka (2008) and Eq. (I-5). Table courtesy of A. Haduch-Sendecka.

**Figures captions**

**Figure 1** Plot of the 'state function' F given by equation (I-3) showing the optimum conditions for plant cell/organ growth. Both coordinates, temperature (T) and acidity or basicity (alkaline medium) pH, are rescaled to unity. The best conditions for maximum growth (green) are found at the saddle point (a stationary point). The simulation parameters used like those for auxin-induced growth – see SI Table 1.

**Figure 2** The (a) isotherms (scaled values indicated) as a function of pH (scaled) of the function F, Eq. (I-3); (b) constant pH curves as a function of temperature (scaled). Simulation parameters: $\alpha = 1.7$ and $\beta = 3.52$ (SI Table 1).

**Figure 3** Solutions (state diagrams) for $\alpha$ and $\beta$ exponents in T and pH coordinates (rescaled). The lines of (a, b, c) constant $\alpha$- and (a', b', c') $\beta$-values in (T, pH) coordinates are indicated in the charts. Simulation data are presented in the figures legends. Green colours – negative values, blue – positive.

**Figure 4** Critical behaviour at STs that take place in the cell wall (pH, T – variables). (a) A "kink" (discontinuity in the first derivative) at the critical temperature $T = T_c$. (b) a "log-divergent" ("$\lambda$" – type) solution at the critical pH = $pH_c$. Control parameters: reduced temperature ($\tau$) and reduced pH ($\pi$). Simulation parameters (a) $\beta = 1.93$ and (b) $\alpha = 1.7$ (SI Table 1).

**Figure 5** Critical behaviour at STs that take place in the cell wall (for $\mu$, T – variables). (a) A "log-divergent" ("$\lambda$" – type) solution at $T = T_c$. (b – c) Solution at the critical value of the chemical potential $\mu = \mu_c$; (b) "acid growth", (c) "auxin growth" (SI Table 1). Control parameters: reduced temperature ($\tau$) and chemical potential ($\mu$). Simulation parameters (a) $\alpha = 3.52$, $\beta = 1.93$; (b) $\alpha = 1.87$, $\beta = 3.17$; (c) $\alpha = 1.7$ and $\beta = 3.52$ (SI Table 1).

**Figure 6** Experimental data for cell volume expansion of maize coleoptile segment as a function of time (fusicoccin applied at $\tau_1$). Cross-correlation volume – pH intensity as a function of time delay; retardation is negligible (left inset). pH measured in experiment – polynomial splines indicated (right inset).

**Figure 7 (a)** Growth rate plot obtained for Eq. (II-1) where pH was assumed to be inversely proportional to time (Lüthen et al. (1990), Fig. 6). Simulation data: $\alpha = 3$, $\beta = 2$; arbitrary units. **(b)** Typical growth curve obtained from Eq. (II-2) – all growth phases are clearly visible: initial phase of slow growth, accelerated growth ("inflation phase"), quasi-linear phase, deceleration and saturation (Fogg, 1975). See also Fig. 8 for parameterisation of Eq. (II-2) with respect to pH (8a), temperature (8b) and turgor pressure (8c).

**Figure 8** Volumetric growth calculated from Eq. (II-2). **(a)** At constant pH. P – Y = 0.3 MPa, T = 0.6 (~ 25 °C); pH-scaled values (0.2 – 0.7, increment 0.1) are shown in the plot (T and pH scaled to [0, 1] interval for all charts (see text)). Note the linear character of the solution of Eq. (II-2) for pH = const, and the maximum expansion for the normalised value of pH ~ 0.3 – 0.4 (note that the slope of the volumetric extension does *not* increase sequentially with decreasing pH). **(b)** For changing temperature T. Simulation parameters: P – Y = 0.3 MPa, T = 0.2, 0.4, 0.6, 0.8 (legend); pH changes like ~ 1/t. Note the typical sigmoid-like curve solution of Eq. (II-2) over a [0, 1] time interval

corresponding to the rapid (~ 1/t) pH drop. Maximum extension (rate) is observed approximately for T = 0.6 (~ 25 °C) and decreases at both higher and lower temperatures as is expected. **(c)** For changing effective turgor pressure P – Y. Simulation parameters: P – Y = 0.1, 0.2, 0.3, 0.4, 0.5 MPa (legend); T = 0.6 (~ 25 °C). Note the sigmoid-like growth curve for a rapidly decreasing pH. The growth curves increase with pressure in sequential order. Simulation parameters that are common for all figures: α = 3, β = 2, $c_{pH}$ = $c_T$ = 1, $\Phi_0$ = $10^{-6}$ [1/MPa × s].

**Figure 9** Effect of an external decrease of pH on the accelerated phase ("inflation phase") of the expanding volume of maize coleoptiles as obtained by pH drops induced by $10^{-5}$ M IAA (A) and $10^{-6}$ M FC (B) in an experiment conducted by Lüthen et al. (1990). Squares and triangles – converted (recalculated) separately (point by point), with the help of Eq. (II-2), from the digitized data presented in Fig. 6 by Lüthen et al. (1990); dashed curves – sigmoid curve (Boltzmann) fit (Microcal Origin). Conversion data: α = 3 and β = 2; $c_T$ = $c_{pH}$ = 1. Simulation parameters: T = 0.62 (~ 26 °C) and P – Y = 0.3 MPa. Note that both fits attain enormously high determination coefficients, which demonstrate that the model is positively verified by experiment (the output is generated based on the pH data and wall properties, see text).

**Figure 10** Plant cell expansion recurring model – schematic view (though, in general, it should be solved self-consistently). A. Initial state: the chemical potential μ ($μ_0$ – reference potential) equilibrated in both compartments; turgor pressure equals P; cell volume equals V. B. Energy (E = hv) consuming proton pump driven by hydrolysis of ATP transport protons (ions) against the electrical gradient. C. Chemical potential difference is established between the compartments – the system is far from equilibrium; turgor pressure equals P; cell volume V. C. Synthesis of cellulose at the plasma membrane and of pectin and hemicelluloses components with a Golgi apparatus deposit a layer on the inside of the existing cell wall causing an increase of the cytoplasm pressure by the infinitesimal value δP: P + δP(t), see Pietruszka and Haduch-Sendecka (2015a), though cell volume still remains V. D. New wall synthesis. Elevated turgor pressure P by δP(t) amount (Benkert et al., 1997) produces force F acting on the wall (energy stored PδV). Chemical potential starts to drop due to the overall particle loss in the system. E. Increasing turgor pressure releases energy to extend the existing cell wall chamber by δl; chemical potential further decreases before the next cycle (arrow). After the completed growth cycle, turgor pressure diminishes by the δP(t) value due to leakage or the other mechanisms illustrated in the text. Another growth cycle begins (arrow), however, with a new initial value of P equal to P – δP(t), V equal to V + δV and μ equal to μ – δμ (exaggerated). Legend. Left compartment: plasma membrane (PM) and cell wall. Right compartment: cytoplasm. Grey colour – PM, blue colour – $H^+$-"Fermi sea", green colour – new cell wall.

**Table 1**

| clade | Monocotyledones | | Dicotyledones | |
|---|---|---|---|---|
| species | *Zea mays* L. | *Triticum vulgare* Vill. | *Phaseolus vulgaris* L. | *Cucurbita pepo* L. |
| α | 2.88 ± 0.28 | 2.91 ± 0.49 | 2.58 ± 0.57 | 5.5 ± 1.1 |
| β | 4.77 ± 0.51 | 3.88 ± 0.71 | 4.3 ± 1.1 | 11.1 ± 2.4 |

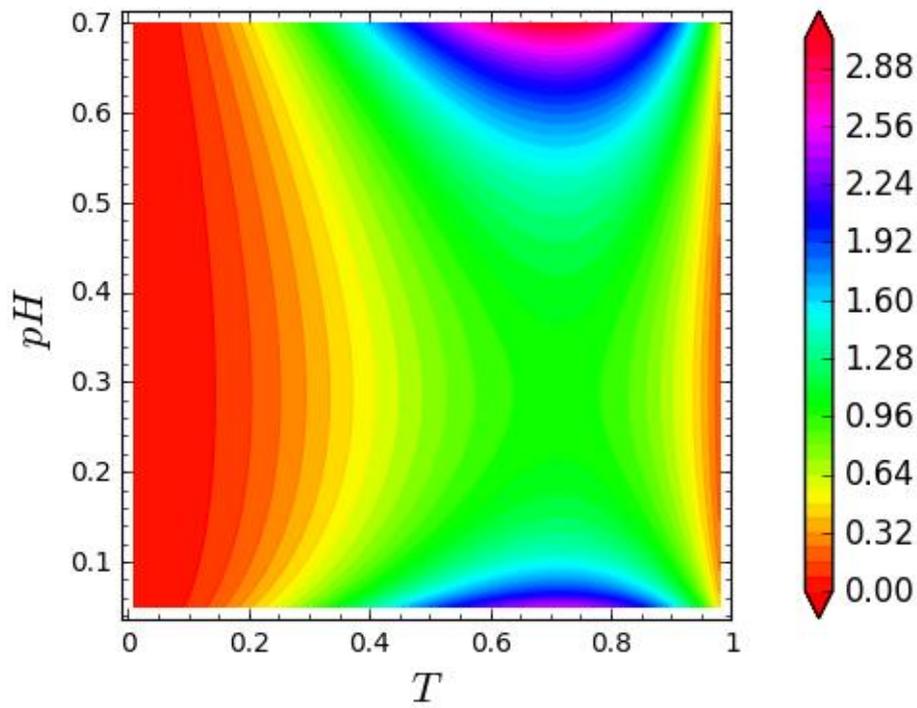

**Figure 1**

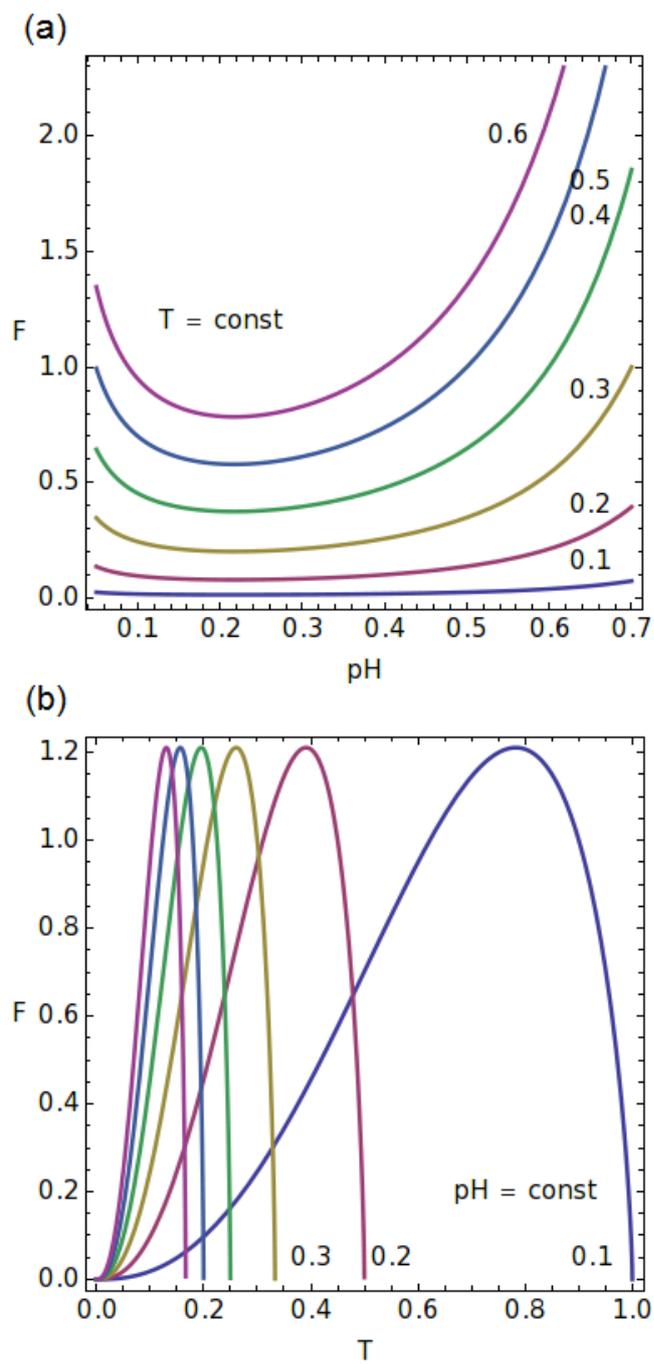

**Figure 2**

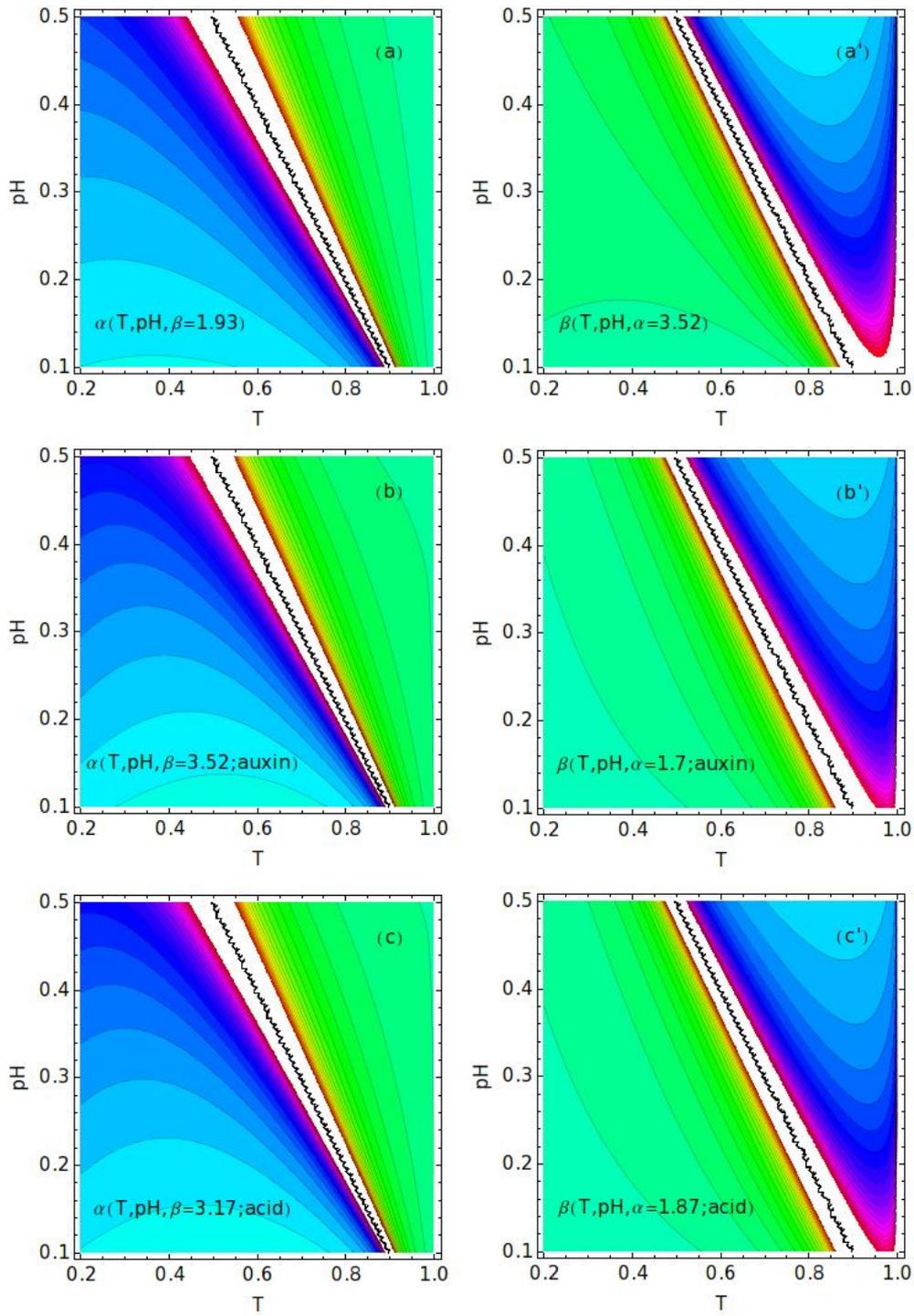

**Figure 3**

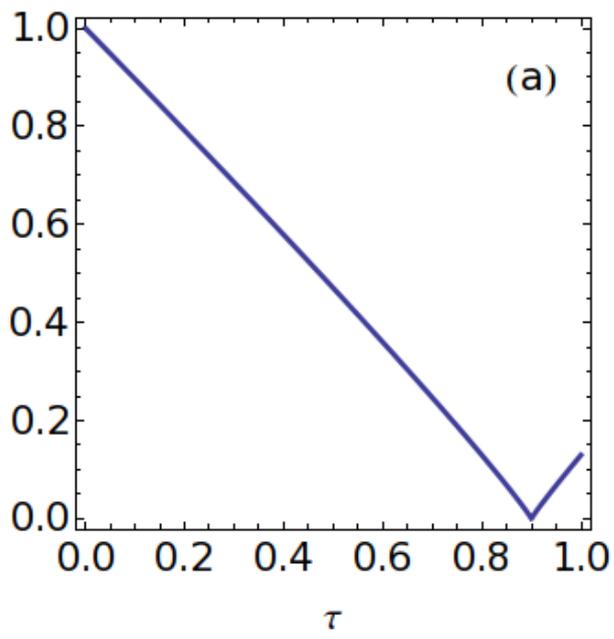

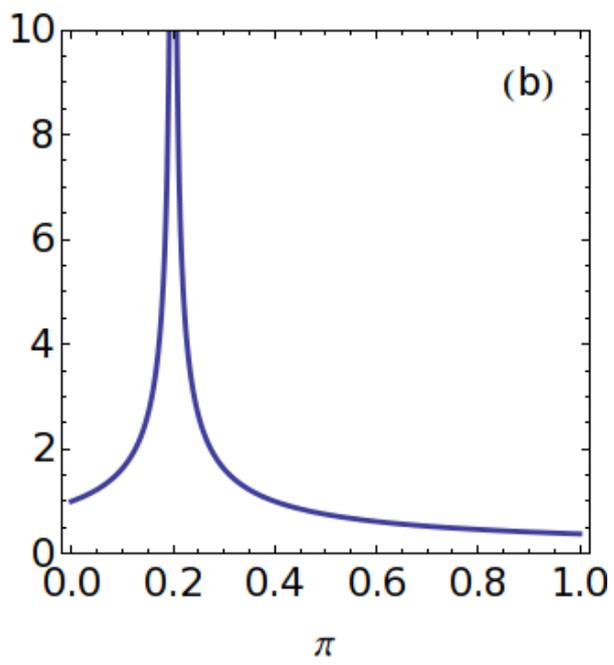

**Figure 4**

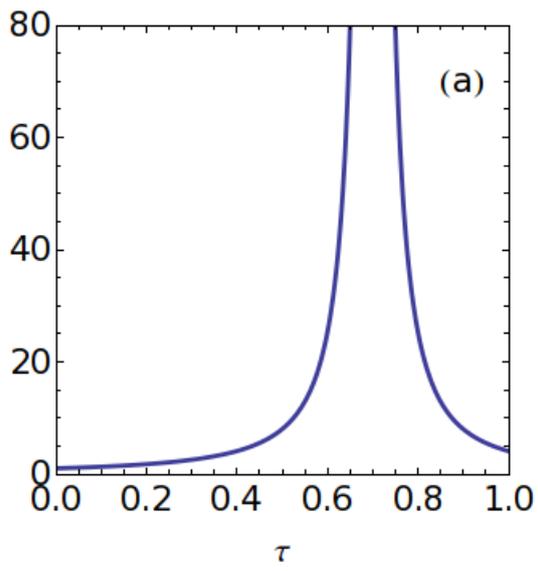

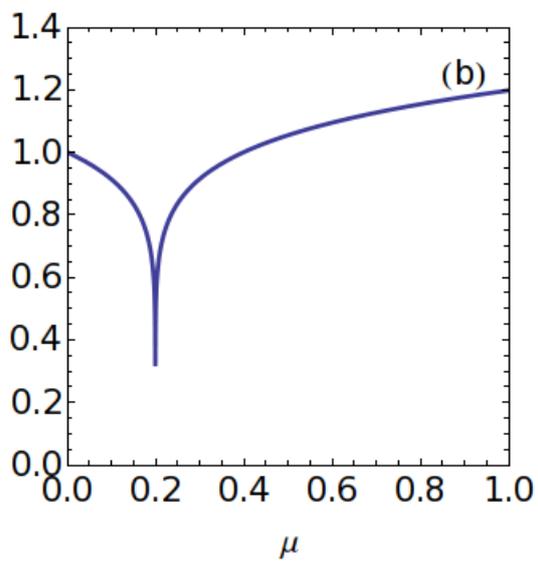

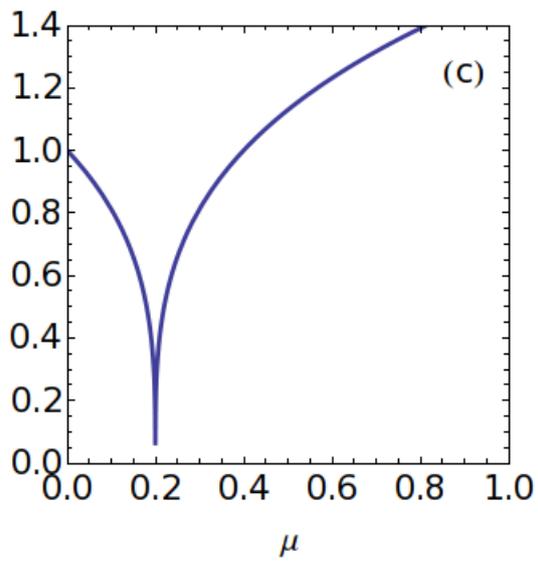

**Figure 5**

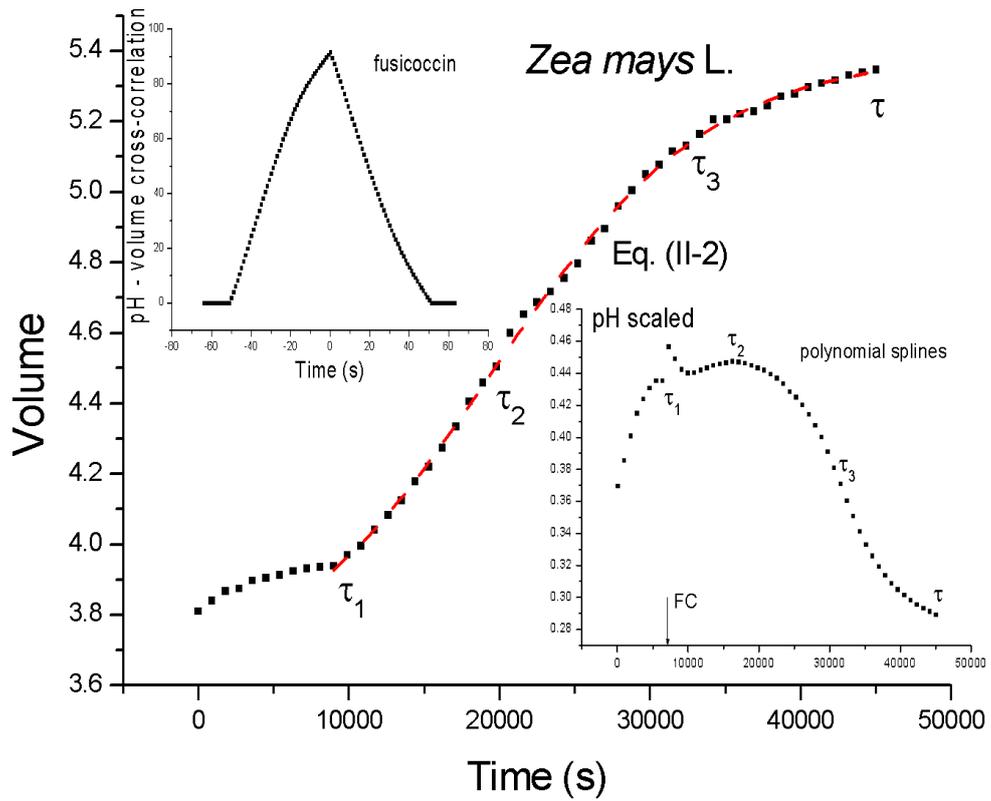

Figure 6

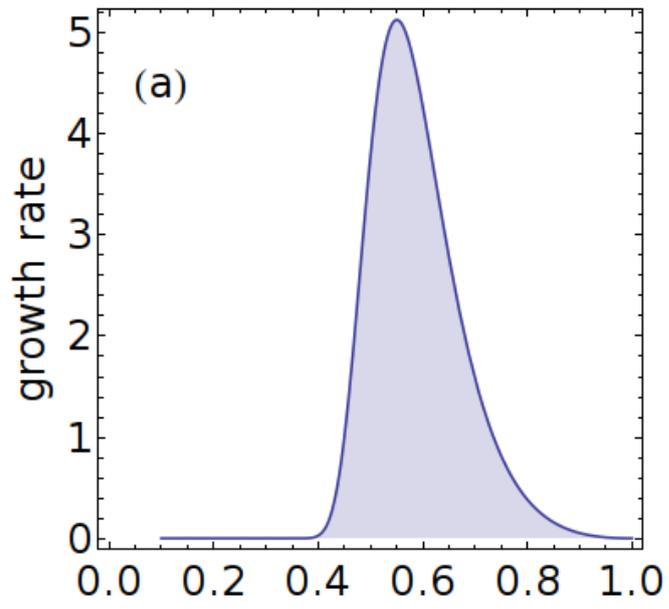

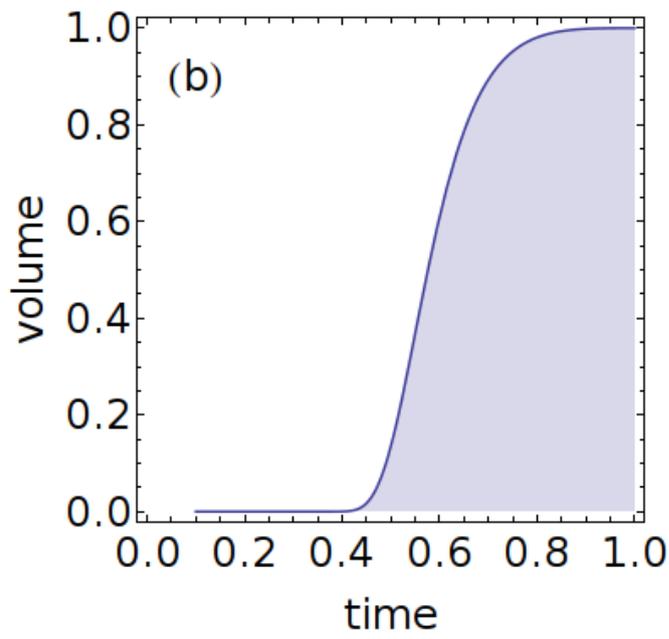

**Figure 7**

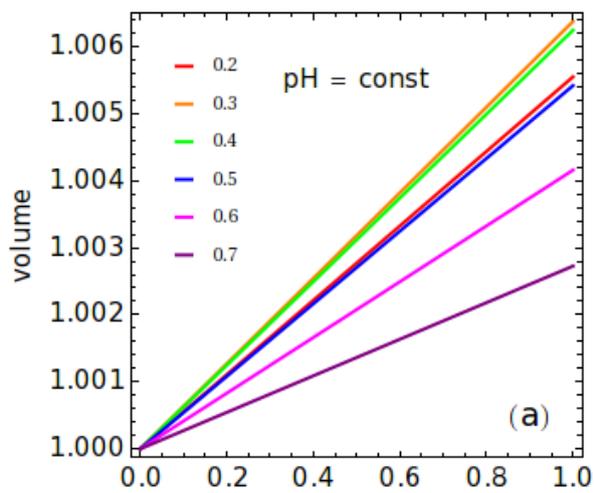

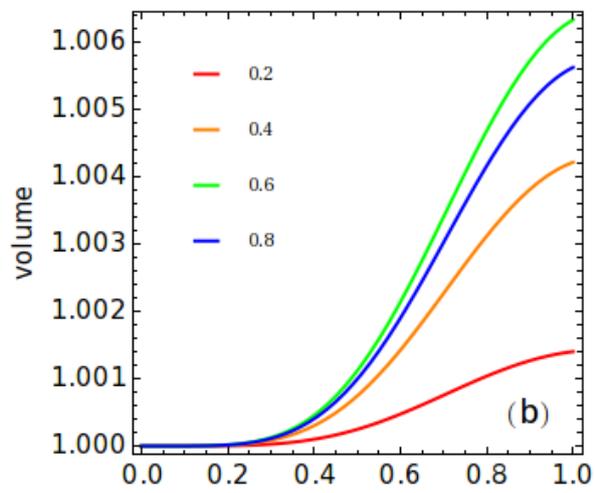

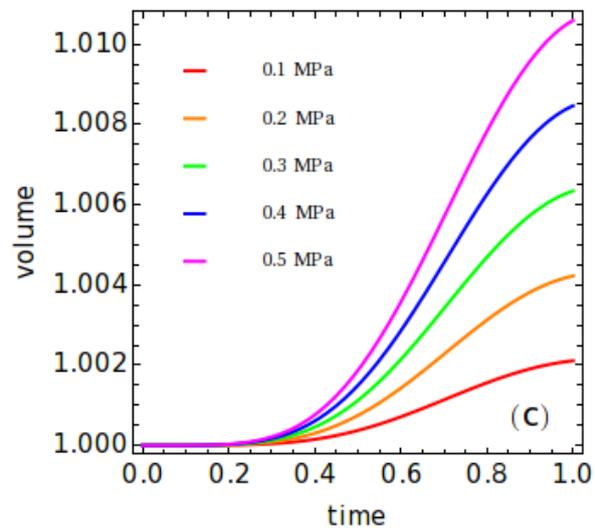

**Figure 8**

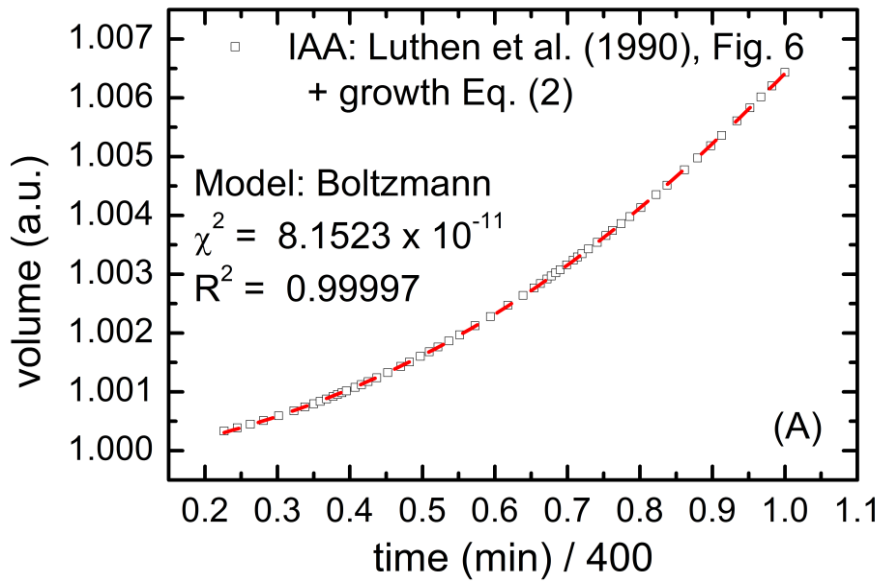

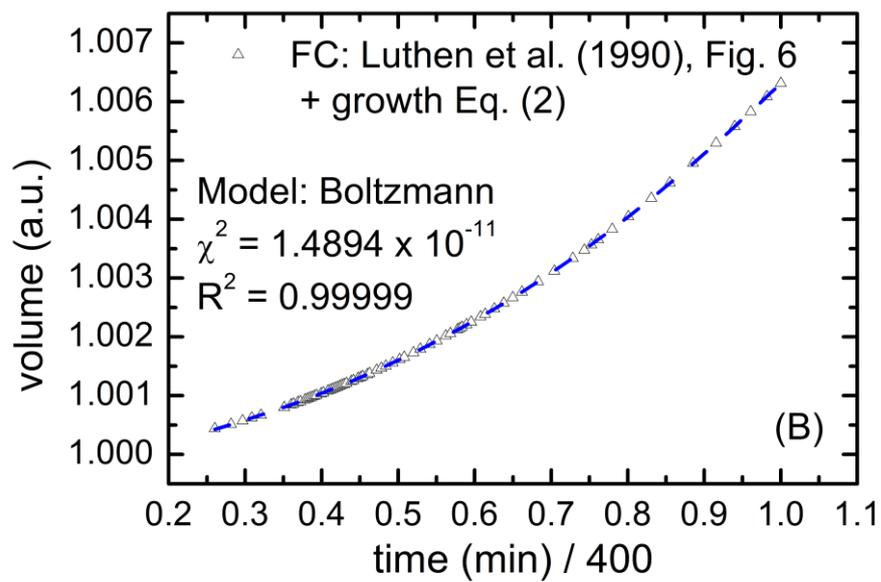

**Figure 9**

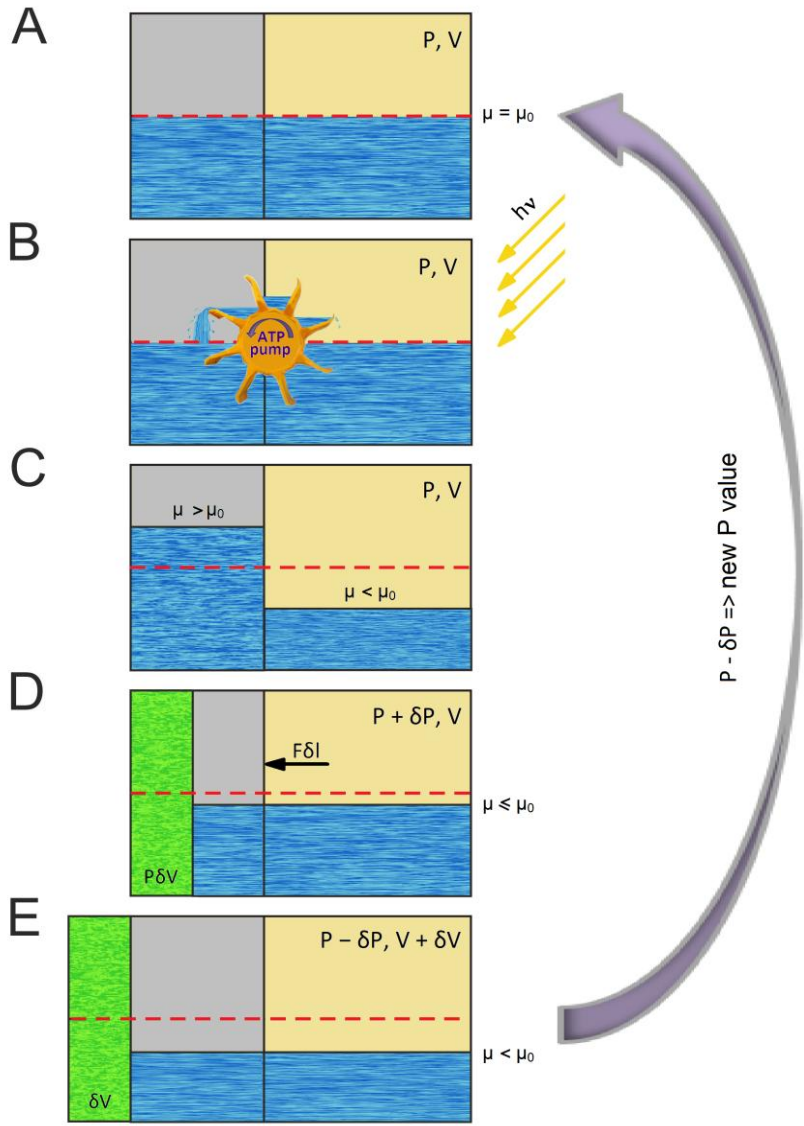

**Figure 10**